\documentclass[12pt, a4paper]{article}

\usepackage[noblocks]{authblk}

\usepackage{amsmath}
\usepackage{graphicx}
\usepackage{xcolor,colortbl}

\definecolor{Gray}{gray}{0.85}
\definecolor{LightCyan}{rgb}{0.88,1,1}

\usepackage{psfrag,epsf}
\usepackage{enumerate}
\usepackage{natbib}
\usepackage{multibib}
\usepackage{url} 

\newcommand{\ech}{\color{black}}

\usepackage{wrapfig}

\usepackage{graphicx}
\usepackage{tikz}
\usetikzlibrary{fit, positioning, arrows, automata}
\usetikzlibrary{shapes, shadows, arrows, positioning, graphs}

\usepackage{amssymb, verbatim, color}
\usepackage{bm,lscape}
\usepackage{wasysym}
\usepackage{bbm}
\RequirePackage[colorlinks, citecolor=blue, urlcolor=blue]{hyperref}
\usepackage{theorem}{}
\usepackage{subcaption}
\usepackage{algorithm}
\usepackage[noend]{algpseudocode}
\usepackage{float}
\usepackage{placeins}

\def\bSig\mathbf{\Sigma}

\definecolor{darkblue}{rgb}{0.0, 0.0, 0.55}

\newcommand{\beq}{\begin{equation}}
	\newcommand{\eeq}{\end{equation}}

\theoremstyle{plain}

\allowdisplaybreaks

\providecommand{\keywords}[1]
{
  \small	
  \textbf{\textit{Keywords---}} #1
}

\title{Entropy regularization in probabilistic clustering}

\author[1]{Beatrice Franzolini}
\author[2]{Giovanni Rebaudo}
\affil[1]{Bocconi University, Milan, IT \href{mailto:franzolini@pm.me}{franzolini@pm.me}}

\affil[2]{University of Turin \& Collegio Carlo Alberto, Turin, IT \href{mailto:giovanni.rebaudo@unito.it}{giovanni.rebaudo@unito.it}}

\date{}

\begin{document}

\maketitle

\begin{abstract}
Bayesian nonparametric mixture models are widely used to cluster observations. 
However, one major drawback of the approach is that the estimated partition often presents unbalanced clusters' frequencies with only a few dominating clusters and a large number of sparsely-populated ones.
This feature translates into results that are often uninterpretable unless we accept to ignore a relevant number of observations and clusters.
Interpreting the posterior distribution as penalized likelihood, we show how the unbalance can be explained as a direct consequence of the cost functions involved in estimating the partition.
In light of our findings, we propose a novel Bayesian estimator of the clustering configuration.
The proposed estimator is equivalent to a post-processing procedure that reduces the number of sparsely-populated clusters and enhances interpretability. 
The procedure takes the form of entropy-regularization of the Bayesian estimate. 
While being computationally convenient with respect to alternative strategies, it is also theoretically justified as a correction to the Bayesian loss function used for point estimation and, as such, can be applied to any posterior distribution of clusters, regardless of the specific model used.
\end{abstract}

\keywords{Dirichlet process, Loss functions, Mixture models, Unbalanced clusters, Random partition}

\section{Introduction}
\label{sec:1}

Clustering methods are used to detect patterns by partitioning observations into different groups. 
What are desirable characteristics of clusters depends on the specific applied problem at hand \citep[see e.g.,][]{hennig2015true}.
Nonetheless, clustering methods are typically motivated by the idea that observations are more similar within the same cluster than across clusters (accordingly to a certain definition of similarity).
Clustering has been proven useful in a large variety of fields including but not limited to image processing, bio-medicine, marketing, and natural language processing.
Clustering methods are used not only to detect sub-groups of subjects, but also for dimensionality reduction \citep{blei2003latent,petrone2009hybrid}, outlier-detection \citep{shotwell2011bayesian,ngan2015outlier,franzolini2023model}, testing for distributional homogeneity \citep{rodriguez2008nested,camerlenghi2019latent,denti2023common,beraha2021semi,balocchi2021clustering,lijoi2023flexible}, and data pre-processing \citep{zhang2006clustering}. 

Among clustering techniques, we can distinguish two main classes: model-based and non-model-based. 
The former methods are built on some assumptions about the sampling mechanism generating the observations. 
The latter are algorithmic procedures computing clusters' allocations without using distributional assumptions and they typically maximize a certain dissimilarity between clusters (or a measure of similarity of the points clustered together). 
Contrary to algorithmic clustering techniques, such as k-means or hierarchical clustering, model-based methods allow us to perform inference via rigorous probabilistic assessments, providing a natural way of quantifying uncertainty.
Importantly, when the assumption about the data generating mechanism is coherently extendable to future data \citep[for example, in infinite exchangeable models, such as the Dirichlet process mixture][]{ferguson1983bayesian,lo1984class}, model-based clustering produces coherent predictions for any number of future observations, based on the available past observations. 
More precisely, by \emph{coherently extendable} we mean preserving \emph{Kolmogorov consistency}, sometimes also called \emph{marginal invariance} \citep{dahl2017random} or \emph{projectivity} \citep{betancourt2022random, rebaudo2023graph}, meaning that the marginal distribution of a sample of size $n$, obtained by marginalizing out the clustering configuration, is equal to the restriction of the distribution of larger samples of size $N>n$.
Thus, their statistical power is not limited to providing a summary of the observed data, as it happens with algorithmic non-model-based techniques. 

Typically, model-based clustering frameworks are equivalent to the assumption that the observations $y_1,\ldots y_n$ are extracted from an infinite population following a mixture
\begin{equation}\label{eq:mix}
y_i\overset{iid}{\sim} \sum_{h=1}^K w_h \, k(\cdot; \theta_h) \qquad \text{for } i = 1,\ldots, n,\,n+1,\ldots,
\end{equation}
where the mixture components  $k(\cdot; \theta_h)$ are probability kernels to be interpreted as distributions of distinct clusters in the infinite population, $(w_h,\,\theta_h)_{h=1}^K$ are unknown parameters that determine the relative proportion and the shape of such population clusters, and $K$ is the total number of clusters in the infinite population. 
$K$ can be either a fixed value or an unknown parameter. In the following, we focus on those models under which either $K$ is unknown or $K=+\infty$. 
Assuming a fixed finite value for $K$ is often restrictive because it limits the flexibility of the assumption in \eqref{eq:mix}, and should be avoided unless we have strong information/preference about an upper bound in the number of clusters. 
Indeed, an important and typically unknown parameter is the number of clusters $K_n$  in the observed sample, i.e., the number of occupied components in the mixture in \eqref{eq:mix}. 
Obviously, $K_n \leq \min(K,n) $.
For this reason, in a framework in which $n$ is let to vary, $K$ is typically either fixed to $+\infty$ \citep[e.g., in Dirichlet process mixtures,][]{ferguson1983bayesian,lo1984class} or it is estimated from the data \citep[e.g., mixtures of finite mixtures, see][]{nobile1994bayesian,miller2018mixture,argiento2022infinity}.  

One limitation typically encountered in model-based clustering is that the clustering point estimate presents highly unbalanced cluster frequencies. 
Especially when the number of mixture components is not arbitrarily fixed to a finite small number, the estimated partition tends to include only a few dominating clusters and a large number of sparsely-populated ones. 
This problem is well-known in Bayesian discrete mixtures such as Dirichlet process mixtures, Pitman-Yor process mixtures, and mixture of finite mixtures. 
This feature is undesirable and poses important problems in terms of interpretability. 
High unbalance in the cluster frequencies typically forces us to disregard all observations assigned to small clusters and just interpret the more-populated ones, for which enough observations are available. 
However, the number of small clusters is often not negligible, so the total number of ignored observations in the interpretation of the cluster is not negligible as well. 
Disregarding observations assigned to small clusters when it comes to model-based clustering is not justified, especially in light of the fact that the unbalance in clusters' frequencies ultimately appears as a feature of the method and not of the specific data analyzed.    

The unbalance of the cluster frequencies can easily be explained as the result of the interaction of the \emph{rich that get richer} property and the unbounded number of clusters in Bayesian mixture models. See \cite{lee2022rich} for a recent detailed discussion on the topic.
The Bayesian learning mechanism of the \emph{rich that get richer}, as the number of observations increases, increases the probability of observing members of clusters that have already been observed. 
At the same time, both in infinite mixture models and in mixture of finite mixtures with a prior on the number of components that assigns positive probability on an infinite set, the probability of observing a new cluster is always positive for any $n$ number of observations already allocated.
Thus, when new observations are collected the induced learning mechanism tends to both repopulate large existing clusters (due to the \emph{rich that get richer} property) and to create new small clusters (due to the fact that the probability of observing new clusters is positive).
The interactions between these two properties naturally reduce into unbalanced clusters. 
However, correcting unbalance intervening on one of these two properties is not optimal. 
It requires either fixing a small upper bound for the number of clusters (i.e., $K$) or breaking probabilistic properties of the model as Kolmogorov consistency of the law of the observable variables \citep[see, for instance,][]{wallach2010alternative,lee2022rich}.

Here we propose a correction of the clusters' unbalance that affects neither the Bayesian learning mechanism nor the attractive probabilistic properties of model-based clustering. 
Our proposal is theoretically justified as a correction to the loss function used for the Bayesian estimate that explicitly reflects the loss in which the analyst incurs when the point estimate of the clustering configuration is uninterpretable. 

The content of the paper is organized as follows. Section~\ref{sec:2} presents the study of the cost functions involved in BNP clustering models and explains the presence of noisy and sparsely populated clusters typically observed in the posterior estimates of these models. 
Then, in light of this study, our computationally convenient and theoretically justified solution to reduce the number of sparsely populated clusters is presented in Section~\ref{sec:3} and showcased on simulated and real data, respectively in Sections~\ref{sec:4} and \ref{sec:5}.
The code to reproduce all results in the paper is available at \url{https://github.com/GiovanniRebaudo/ERC}.

\section{Implied costs functions in Bayesian nonparametric clustering}
\label{sec:2}
The main goal of clustering techniques is to estimate a partition of the observed sample, more than the distribution of the whole ideal population in \eqref{eq:mix}. 
The partition that one wants to estimate can be encoded using a sequence of subject-specific labels $(c_1,\ldots,c_n)$ taking value in the set of natural numbers such that $c_i = c_j = c$ if and only if  $y_i$ and $y_j$ belong to the same cluster and follow the same mixture component $k(\cdot; \theta_c)$, i.e. $y_i \mid c_i \overset{ind}{\sim} k(\cdot; \theta_{c_i})$  for ${i = 1,\ldots, n}$.
The indicators $(c_1,\ldots,c_n)$, as just defined, are affected by the label switching problem \citep[see, for instance,][]{stephens2000dealing, mclachlan2019finite, gil2020beta}. 
In the following, we assume them to be encoded in order of appearance. 
This means that $c_1 = 1$, i.e. the first observation $y_1$ always belongs to the \emph{first} cluster. 
Then either $c_2 = c_1 = 1$, if the second observation $y_2$ is clustered together with $y_1$, or $c_2=2$, otherwise, and so on and so forth.
Note that, thanks to exchangeability, we can focus on an arbitrary order of the observations without affecting the joint law of the sample and thus posterior inference of the clustering configuration.
The likelihood for $\bm{c} = (c_1,\ldots,c_n)$ and $\bm{\theta} = (\theta_1,\ldots,\theta_{K_n})$ is
\begin{equation}\label{eq:lik}
\mathcal{L}(\bm{c},\bm{\theta};\bm{y}) = \prod_{c=1}^{K_n} \prod_{i : c_i = c} k(y_i; \theta_{c}).
\end{equation}

When $K_n$ is unknown, the clustering labels in \eqref{eq:lik} cannot be estimated with a standard frequentist approach. 
In fact, when the maximum likelihood estimator (MLE) for \eqref{eq:lik} exists, it coincides with the vector of MLEs $(\hat{\theta}_{1},\ldots, \hat{\theta}_n )$, where each $\hat{\theta}_i$ is obtained considering one observation at a time and the independent models $y_i \sim k(y_i\mid \theta_i)$, for $i = 1,\ldots,n$. 
This result is an immediate consequence of
\[
\max_{(\bm{\theta}, \bm{c})} \sum_{i=1}^{n} \log k(y_i; \theta_{c_i}) \leq \sum_{i=1}^{n} \max_{\theta_{i}} \log k(y_i; \theta_{i})
\]
and when there are no joint constraints among the parameters in $\bm{\theta}$ the equality holds.

Moreover, note that under typical mixture model assumptions for clustering, we have that $\hat{\theta}_{1} \ne \ldots \ne \hat{\theta}_n $. 
For instance, when $k$ is a multivariate Gaussian density and $\theta$ is the pair of mean vector and variance matrix of the Gaussian component, such the MLE entails a number of clusters equal to the number of distinct observed values, that by model's assumptions equals $n$ with probability 1. 
Thus, no information on clusters can ever be gained through MLE and overfitting is unavoidable unless one relies on strong restrictions of the parameter space \citep[cfr.\ also with Theorem 1 and 2 in][where the use of a uniform prior over all possible partitions is considered]{casella2014cluster}. 
In this regard, note that maximizing \eqref{eq:lik} is not the same as computing the nonparametric maximum likelihood estimator \citep{lindsay1995mixture, polyanskiy2020self, saha2020nonparametric} for the mixture model in \eqref{eq:mix}.

Differently, Bayesian models, and in particular Bayesian nonparametric (BNP) models, are largely used for model-based clustering, since priors act as penalties, shrinking the number of distinct clusters.
The vast majority of Bayesian models for clustering rely on a prior for $\bm{c}$ and $K_n$ defined through an exchangeable partition probability function (EPPF) \citep[see,][]{pitman1996some} and, independently, a prior $P$ is used for the unique values $(\theta_1,\ldots,\theta_{K_n})$. 
Recall that an EPPF characterizes the distribution of an exchangeable partition, with $\text{EPPF}(n_{1},\ldots,n_{K_{n}})$ being the probability of observing a particular (unordered) partition of $n$ observations into $K_{n}$ subsets of cardinalities $\{n_{1},\ldots,n_{K_{n}}\}$.

Therefore, the corresponding posterior distribution is 
\begin{equation}\label{eq:posterior}
p(K_n, \bm{c}, \bm{\theta} \mid \bm{y}) \propto \prod_{c=1}^{K_n} \prod_{i : c_i = c} k(y_i; \theta_{c}) \times \text{EPPF}(n_1,\ldots,n_{K_n}) \times P(d \bm{\theta}),
\end{equation}
which can be equivalently represented as the cost function $- \log(p(K_n, \bm{c}, \bm{\theta} \mid \bm{y})) $, i.e.
\begin{equation*}
C(K_n, \bm{c}, \bm{\theta}; \bm{y}) = C_{\text{lik}}(K_n, \bm{c}, \bm{\theta}; \bm{y}) + \,C_{\text{part}}(K_n, \bm{c}; \alpha) + \,C_{\text{base}}(K_n, \bm{\theta}),
\end{equation*}
which is the sum of three terms, that in the following are named respectively likelihood cost, partition cost, and base cost. 

As already mentioned, the minimum likelihood cost 
\begin{equation*}
C_{\text{lik}}(K_n, \bm{c}, \bm{\theta}; \bm{y}) = -\sum_{c=1}^{K_n}\sum_{i:c_i = c}^n \log\, k(y_i; \theta_{c})
\end{equation*}
typically corresponds to $K_n$  equal to the number of distinct observed values. 
The remaining two costs are those defined by the prior of the model and their marginal behavior is described here below.
Clearly, any inference result has to be derived based on the whole posterior distribution in \eqref{eq:posterior}, which is the result of the joint, and not marginal, effects of all three costs.
Nonetheless considering one cost at a time allows us to gain insights regarding the estimation procedure and the frequentist penalties induced by the prior. 
\subsection{Base cost}
\begin{figure}[!t]
	\begin{subfigure}[t]{0.49\textwidth}
		\centering
		\includegraphics[width=\textwidth]{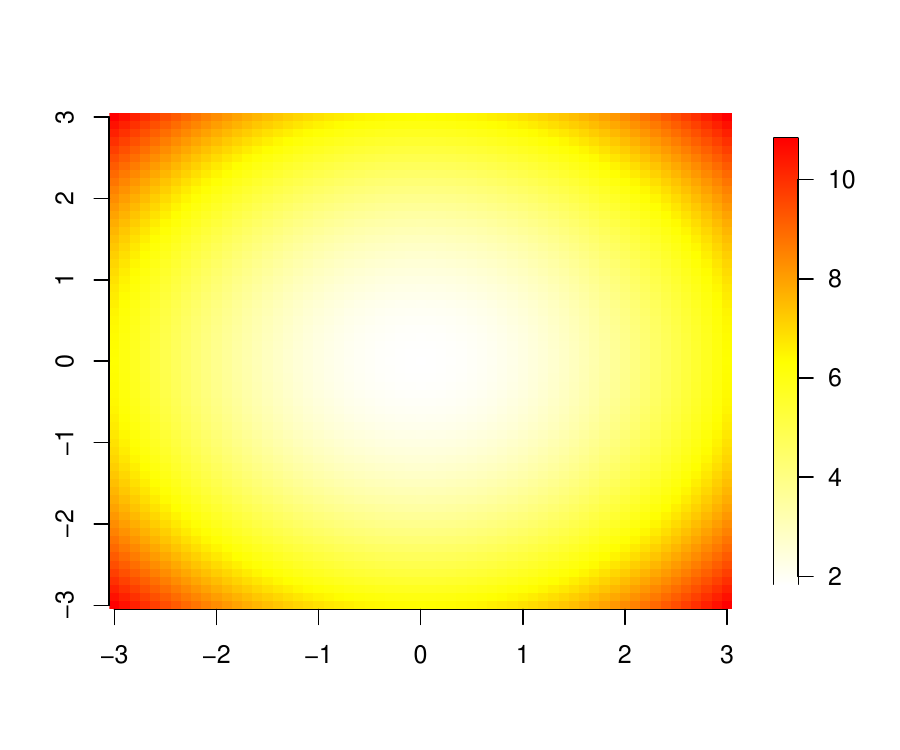}
		\caption{\label{fig:baseA}Base cost for $(\theta_1, \theta_2)$ with $\sigma^2 = 1$}
	\end{subfigure}	
	\hfill
	\begin{subfigure}[t]{0.49\textwidth}
		\centering
		\includegraphics[width=\textwidth]{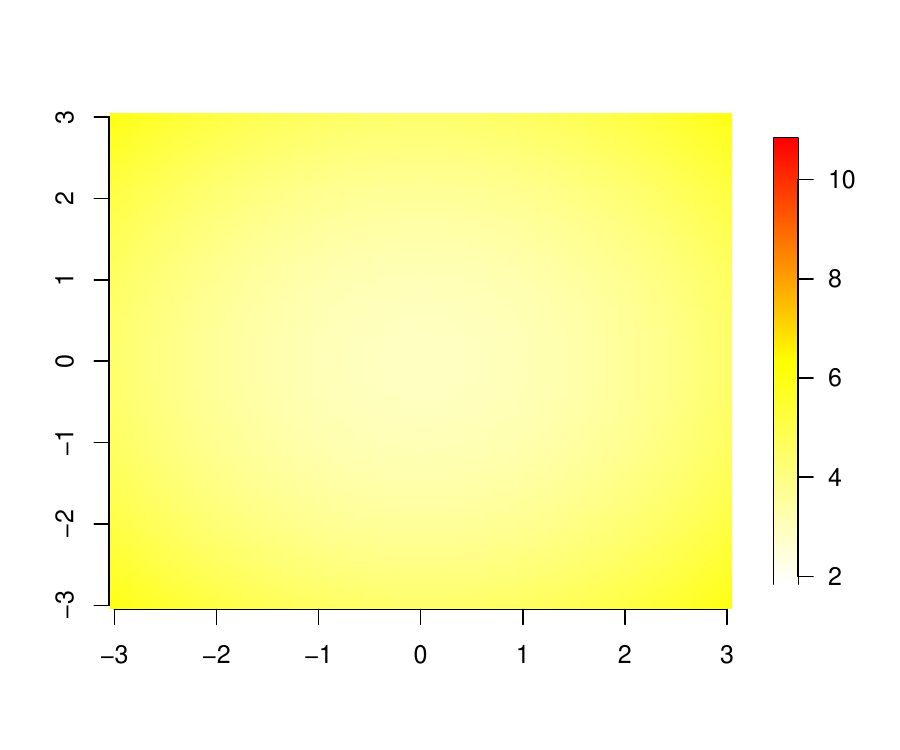}
		\caption{\label{fig:baseB}Base cost for $(\theta_1, \theta_2)$ with $\sigma^2 = 3$}
	\end{subfigure}
 \\[5ex]
    \begin{subfigure}[b]{0.45\textwidth}
		\centering
		\includegraphics[width=\textwidth]{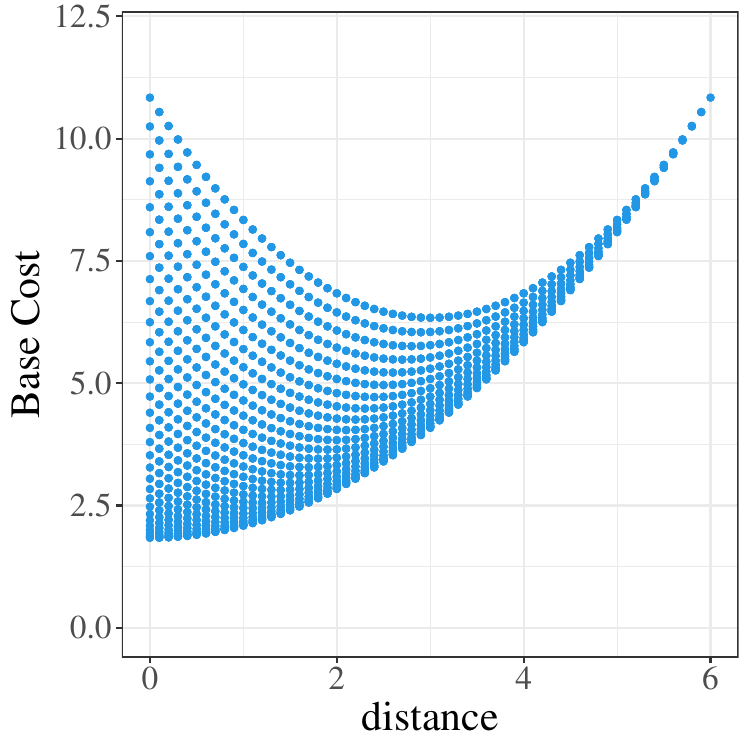}
		\caption{\label{fig:baseC}Base cost by distance between $\theta_1$ and $\theta_2$ with $\sigma^2 = 1$. 
  Each point corresponds to a vector $(\theta_1, \theta_2)$.}
	\end{subfigure}	
	\hfill
	\begin{subfigure}[b]{0.45\textwidth}
		\centering
		\includegraphics[width=\textwidth]{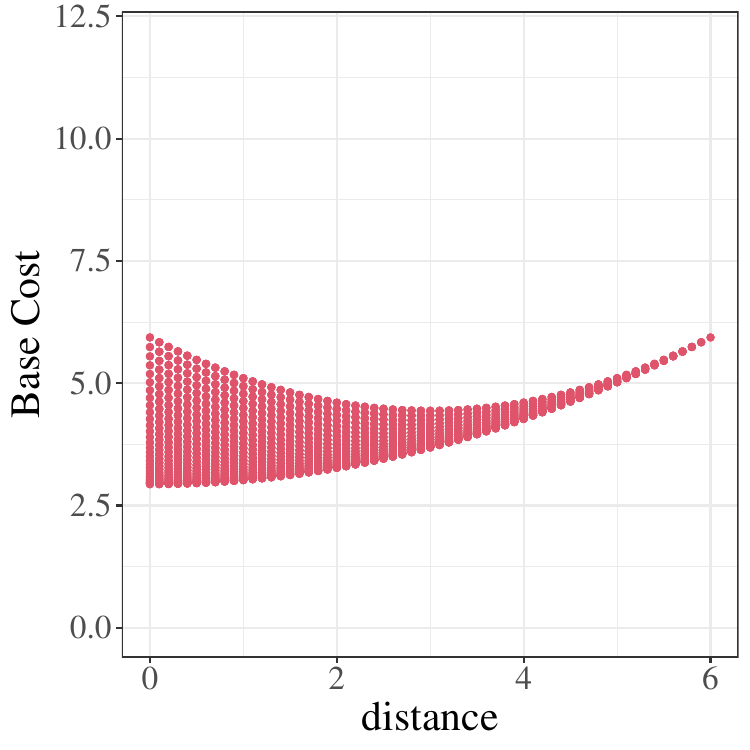}
		\caption{\label{fig:baseD}Base cost by distance between $\theta_1$ and $\theta_2$ with $\sigma^2 = 3$. 
  Each point corresponds to a vector $(\theta_1, \theta_2)$.}
	\end{subfigure}	
	\caption{ \label{fig:basecost} $K_n = 2$, bivariate normals with $\sigma^2 = 1$ and $\sigma^2=3$}
\end{figure}

A lot of attention in the literature has been devoted to the choice of the EPPF and many alternatives are available \citep[see, for example,][]{antoniak1974mixtures, green2001modelling, lijoi2007controlling, lijoi2010models, deblasi2015gibbs, camerlenghi2018bayesian, miller2018mixture, greve2020spying}, while, except for few cases, mainly within repulsive mixtures \citep{petralia2012repulsive,xu2016bayesian,bianchini2020determinantal,xie2020bayesian,beraha2022mcmc}, the role of the base cost appears partially overlooked within the Bayesian methodology literature.

However, when BNP clustering methods are applied in practice, the choice of an appropriate base distribution is known to be crucial.
The most common choice is to use an independent prior on the unique values so that $\theta_c\overset{iid}{\sim}P_0$ and
\begin{equation*}
C_{\text{base}}(K_n, \bm{\theta}) = - \sum_{c=1}^{K_n} \log P_0(d \theta_c),
\end{equation*}
where the variance of the distribution $P_0$ is known to play an important role in the estimation process and, typically, the higher the variance of $P_0$ the lower the number of clusters identified by the posterior \citep[cfr., e.g.][p. 535]{gelman2013bayesian}. 
This phenomenon can be explained by looking at the joint distribution induced by $P_0$ on the unique value.
Higher values of the variance correspond to a joint distribution with a smaller mass around the main diagonal and, therefore, a higher base cost for those vectors $(\theta_1,\ldots,\theta_{K_n})$ whose components are similar, thus ultimately favoring the variability of the unique values and penalizing many overlapping clusters.
Consider for instance the case of $P_0$ set to a univariate normal distribution centered in $\mu$ and with variance $\sigma^2$, we have 
\[
C_{\text{base}}(K_n, \bm{\theta}) = \frac{K_n}{2}\, \log (2 \pi) + \frac{K_n}{2} \log{\sigma^2} + \frac{1}{2} \sum_{c=1}^{K_n} \frac{(\theta_c - \mu)^2}{\sigma^2}.
\]
When the variance is increased from $\sigma^2$ to $\lambda^2$, the base cost increases closer to the $K_n$-dimensional vector $(\mu, \ldots, \mu)$.
More formally, defining the $K_n$-sphere $\bm{\theta} \in \mathbb{R}^{K_n}$ such that $\sum_{c=1}^{K_n} (\theta_c - \mu)^2 = K_n \frac{\log(\lambda^2/\sigma^2)\sigma^2 \lambda^2}{\lambda^2 - \sigma^2}$, we have that  the cost increases for vectors $(\theta_1,\ldots,\theta_{K_n})$ corresponding to points inside the sphere and decreases for those vectors corresponding to points outside the sphere.
This causes also a reduction in the relative cost of those vectors located far from the main diagonal compared to the cost of the vectors closer to the main diagonal.
To clarify this point,  Figure~\ref{fig:basecost} shows the cost function shift caused by an increase in variance from $1$ to $3$ in the case of $K_n = 2$ and $P_0$ univariate normal centered in 0.
In Figure~\ref{fig:basecost}, the number of cluster is fixed to $K_n = 2$ and the cost associated to different clusters' locations $(\theta_1,\theta_2) \in [-3,3]^2$ is considered.
Figures~\ref{fig:baseA} and \ref{fig:baseB} show the base cost in the whole plane $[-3,3]^2$, while Figures~\ref{fig:baseC} and \ref{fig:baseD} show how the base cost changes based on the distance between $\theta_1$ and $\theta_2$, i.e, $|\theta_1-\theta_2|$.
Figures~\ref{fig:baseC} and \ref{fig:baseD} are obtained considering a grid of equally spaced points in the plane $[-3,3]^2$.  
Comparing the two scenarios of variance equal 1 and 3, it is evident as the increase in variance results in a smaller penalization of the distance between cluster locations.
In practice, $P_0$ is usually set to be a continuous scale mixture, where the mixed density is conjugate to the kernel $k$ for computational convenience, while the mixing density is used to increase appropriately the marginal scale of the mixture $P_0$. 

\subsection{Partition cost}
Finally, let us comment on the partition cost $C_{\text{part}}$.
Its behavior is less straightforward and we consider here only two important and widely used cases: Dirichlet process mixtures (DPM) and Pitman-Yor process \citep{pitman1997two} mixtures (PYPM).
With a DPM model, up to an additive constant, we have
\[
C_{\text{part}}(K_n, \bm{c}; \alpha) = -K_n\,\log \,\alpha - \sum_{c=1}^{K_n} \log \Gamma(n_c),
\]
where $\alpha$ is the concentration parameter of the Dirichlet Process. 
The DPM partition cost tends to favor parsimonious values of $K_n$ (with respect to the likelihood cost that in general tends to favor $K_n = n$). 
However, contrary to the base cost, it depends also on clusters' frequencies.
  
\begin{figure}[t]
    \centering
    \includegraphics[width=0.75\textwidth]{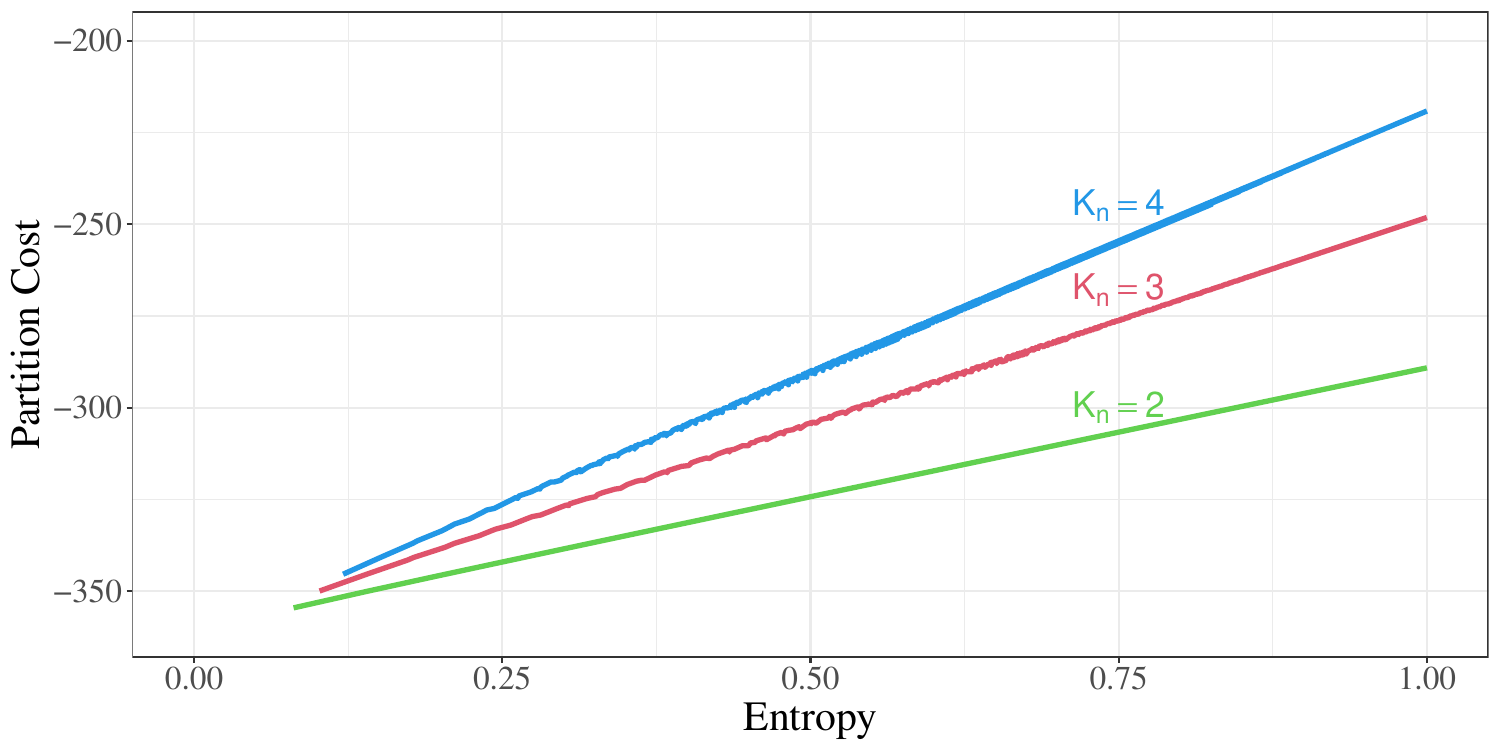}
    \caption{\label{fig:PartCostDPM} 
  Average partition cost as a function of entropy of the partition in a DPM model with $\alpha=1$ for $n=100$ observations clustered into 2 (blue line), 3 (red line), and 4 (green line) clusters.
  Plotted values are obtained analytically: for each possible partition, the value of the entropy and the cost are computed and, then, the cost is averaged across partitions with the same entropy, keeping the number of clusters fixed.}
\end{figure}	

\begin{figure}[t]
    \centering
    \includegraphics[width=0.75\textwidth]{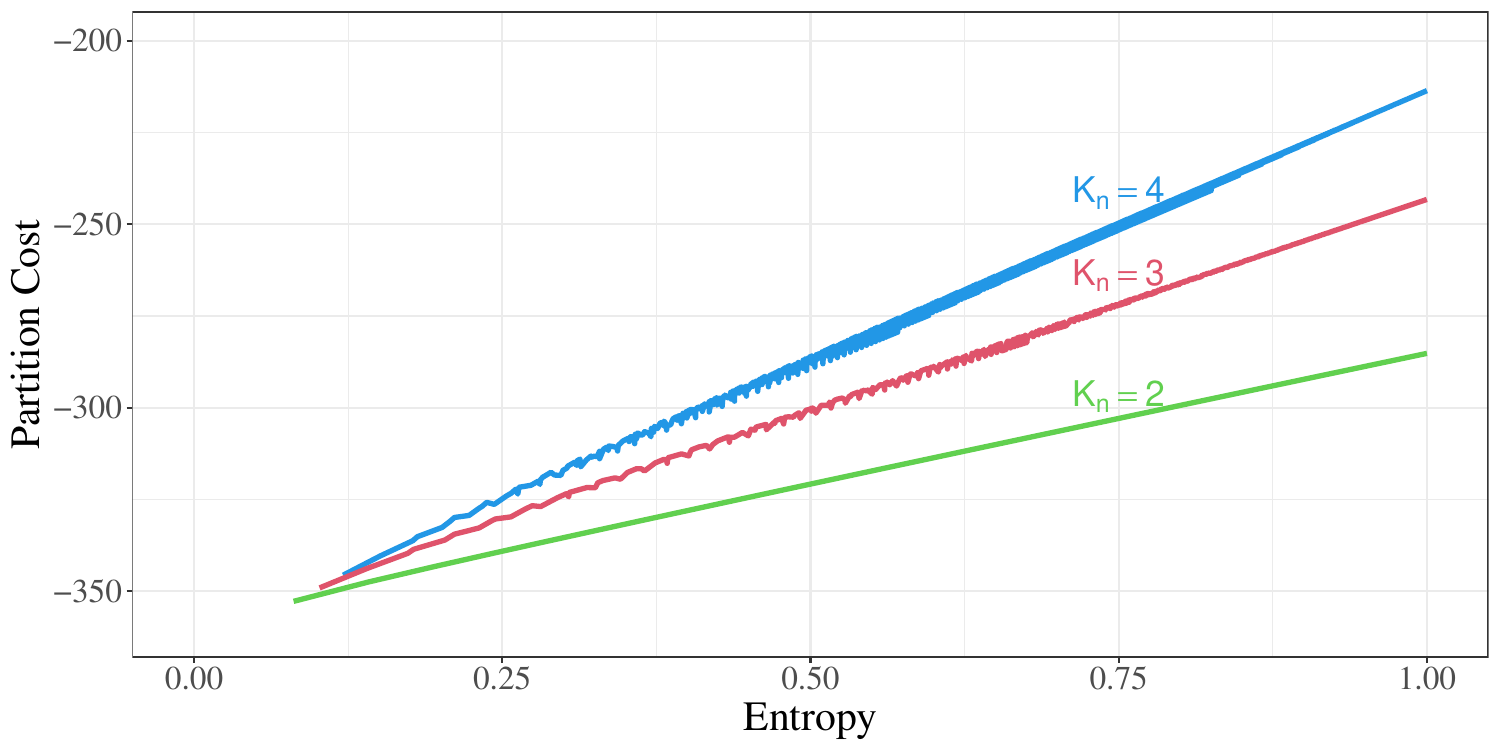}
    \caption{\label{fig:PartCostPYP}
    Average partition cost as a function of entropy of the partition in a PYPM model with $\alpha=1$ and $\sigma = 0.5$ for $n=100$ observations clustered into 2 (blue line), 3 (red line), and 4 (green line) clusters.
    Plotted values are obtained analytically: for each possible partition, the value of the entropy and the cost are computed and, then, the cost is averaged across partitions with the same entropy, keeping the number of clusters fixed.}
\end{figure}
Figure \ref{fig:PartCostDPM} showcases the partition cost of DPM for different values of what we refer henceforth to as the entropy of the frequencies $(n_1,\ldots,n_{K_n})$, i.e. 
\begin{equation*}
 S(n_1,\ldots,n_{K_n}) = - \sum_{c=1}^{K_n}\frac{n_c}{n} \, \log_{K_n} \frac{n_c}{n}.
\end{equation*}
Overall the EPPF acts favoring frequencies $(n_1,\ldots,n_{K_n})$ with low entropy and thus, roughly speaking, higher sample variance of the frequencies. 
However, this feature ultimately results in two distinct effects: one acting on the total number of occupied clusters $K_n$ and another acting on the variance of the clusters' frequencies $(n_1,\ldots,n_{K_n})$. 
Even though these two features both favor a reduced entropy, they entail very different scenarios in terms of estimated clustering structure, especially from an applied and practical point of view.
Penalizing large numbers of clusters is typically desirable in applications because an elevated number of clusters may be difficult to interpret.
However, a partition with few dominating clusters and many sparsely populated clusters is often highly undesirable because it is hard to interpret unless one decides to ignore all the information contained in the small clusters and focus only on the dominating ones.
See also \cite{green2001modelling} for a study of the posterior entropy in the Dirichlet process mixture and \cite{greve2020spying} for more details on entropy in mixtures of finite mixture models.
In the case of a PYPM the partition cost, up to an additive constant, equals 
 \[
 C_{\text{part}}(K_n, \bm{c}; \alpha, \sigma) = - \sum_{c=1}^{K_n} \log (\alpha + \sigma (c - 1) ) -
  \sum_{c=1}^{K_n} \log \Gamma(n_c - \sigma) + 
  K_n \,\log \Gamma(1- \sigma).
 \] 
Despite that the EPPFs are different, Figures \ref{fig:PartCostDPM} and \ref{fig:PartCostPYP} show in both processes a closely similar behavior in terms of entropy penalization. 
This tendency is coherent with the fact that the posterior unbalance of cluster frequencies is typically observed in practice under both models, although they are built on different EPPFs.  

Note that Figures \ref{fig:PartCostDPM} and \ref{fig:PartCostPYP} provide us with insights into the behavior of the EPPFs evaluated (analytically from the aforementioned expressions) in correspondence of different vectors of clusters' frequencies $(n_1,\ldots,n_{K_n})$, i.e., the probability of a specific clustering configuration with unordered frequencies $\{n_1,\ldots,n_{K_n}\}$.  
In particular, they show how the EPPF associates different levels of penalty with different values of entropy.
In this regard is important to stress that the vectors $(n_1,\ldots,n_{K_n})$ are not in a one-to-one correspondence with the partitions, and the number of partitions corresponding to certain frequencies varies across vectors. 
For instance, when $n=100$, there exist $\binom{100}{50, 50} \approx 1.01e+29$ distinct partitions corresponding to the vector of frequencies $(50,50)$ and $\binom{100}{25,25,25,25}\approx1.61e+57$ distinct partitions corresponding to the vector of frequencies $(25,25,25,25)$. 
The number of partitions per different vectors of frequencies, which is not depicted in Figures \ref{fig:PartCostDPM} and \ref{fig:PartCostPYP}, does affect both estimates of \ech marginal quantities, such as the number of clusters $K_n$, as well as point estimates of the clustering configuration that are different from the MAP (maximum a posteriori). 

If we are interested in estimating the number of clusters $K_n$, we should note that the number of possible partitions rapidly changes with $K_n$ accordingly to Stirling numbers of the second kind.
More precisely, the Stirling number of the second kind counts the number of different partitions of $n$ objects into $K_n$ non-empty unordered subsets and can be computed as
\[
\frac{1}{K_n!} \sum_{i=0}^{K_n}(-1)^{K_n- i}{\binom {K_n}{i}}i^{n}.
\]

This information must be combined with the partition cost, as represented in Figures \ref{fig:PartCostDPM}  and \ref{fig:PartCostPYP}, if we are interested in fully understanding the impact of the EPPF on the marginal prior and posterior distributions of $K_n$.
Combining the two features (i.e., the partition cost per each vector of frequencies and the number of partitions per each vector of frequencies) the typical partition cost strongly penalized too many clusters suggested by the likelihood costs, i.e. $K_n=n$, but still favors a small (higher than $1$) number of clusters that adaptively increases with the sample size $n$ \citep[see e.g.,][]{deblasi2015gibbs}.

\section{Regularized-entropy estimator}
\label{sec:3}

Once the posterior distribution $\mathbb{P}(\bm{c} \mid y_{1:n})$ over the space of partitions is obtained, typically thanks to a Markov Chain Monte Carlo algorithm, a point estimate $\hat{\bm{c}}$ of the partition can be obtained accordingly to the decision-theoretic approach of Bayesian analysis.
More precisely, $\hat{\bm{c}}$ is obtained by minimizing the Bayesian risk, i.e, the expected value of a loss function $L(\bm{c}, \hat{\bm{c}})$ with respect to the posterior
\[
\bm{c}^* = \underset{\hat{\bm{c}}}{\text{argmin}}\,\mathbb{E}[L(\bm{c}, \hat{\bm{c}}) \mid y_{1:n}] = \underset{\hat{\bm{c}}}{\text{argmin}} \sum_{{\bm{c}}} L(\bm{c}, \hat{\bm{c}}) \mathbb{P}(\bm{c} \mid y_{1:n}),
\]  
where $L(\bm{c}, \hat{\bm{c}})$ is the loss in which we incur using $\hat{\bm{c}}$ as estimates when the partition takes the value $\bm{c}$.
How to interpret and elicit the loss in practice can change according to the philosophical point of view.
See, for instance, \cite{robert2007bayesian}.
Often in parameter estimation, the loss is interpreted as the cost of choosing $\hat{\bm{c}}$ instead of the ideally optimal parameter value $\bm{c}$ (sometimes interpreted as the \emph{truth}).
In a more subjective Bayesian framework, it can be interpreted, together with the model and prior, in terms of the preferences implied on the possible parameter values $\bm{c}$ via the Bayesian risk \citep{savage1972foundations}.
Finally, also in a more frequentist framework, the loss can be chosen in terms of the implied properties of the estimator $\hat{\bm{c}}$ of the unknown true parameter.

Despite the different philosophical justifications, rarely, in applied Bayesian clustering analysis, a 0-1 loss function and the resulting MAP estimator are employed due to the large support of the posterior and the fact that the 0-1 loss function does not reflect different levels of distance between two non-coinciding partitions. 
Widely used alternatives in applications are Binder loss \citep{binder1978bayesian} or variation of information (VI) loss \citep[see,][]{meila2007comparing, wade2018bayesian, dahl2022search}.
 \begin{algorithm}[t]
	\caption{\label{alg}Entropy-regularized estimate}
 \vspace{0.3 cm}
	\textbf{Inputs}: chain of partitions $\{\bm{c}_m, m=1,\ldots,M\}$ sampled from the posterior, $\lambda$\\
	\textbf{Output}: point estimate $\bm{c}^*$
 \vspace{0.3 cm}
	\begin{algorithmic}[1]
		\State Compute $S(\bm{c}_m)$ for $m =1,\ldots,M$\vspace{0.2cm}
		\State Compute $w_m = \exp \{\lambda S(\bm{c}_m)\}$ for $m =1,\ldots,M$\vspace{0.2cm}
		\State $\bar w_m \gets w_m / \sum_m w_m$ for $m =1,\ldots,M$\vspace{0.2cm}
		\State Generate $\{\tilde {\bm{c}}_m, m =1,\ldots,M\}$, sampling with replacement from $\{\bm{c}_1,\ldots,\bm{c}_M\}$ with prob. $\{\bar w_m, m=1,\ldots,M\}$\vspace{0.2cm}
		\State $\bm{c}^* \gets \text{argmin} \sum_{m = 1 }^M L(\tilde {\bm{c}}_m, \hat{\bm{c}})$\vspace{0.2cm}
	\end{algorithmic}
\end{algorithm}

We have already stressed how a large presence of noisy clusters is typically undesirable in practice and we claim that this aspect should be reflected in the loss function used for point estimation so that the loss of each partition is proportional to its entropy. 
To do so, consider any possible loss function $L(\bm{c}, \hat{\bm{c}})$ one would like to use to derive the estimate, we can define a new loss function, that we named entropy-regularized, as 
\[
\bar L(\bm{c}, \hat{\bm{c}}) = \exp\{\lambda S({\bm{c}})\} L(\bm{c}, \hat{\bm{c}}),
\]
where, with a little abuse of notation w.r.t.\ the previous section, $S({\bm{c}})$ is the entropy of the partition identified by ${\bm{c}}$ and $\lambda \in \mathbb{R}$. 
Recall that the base of the logarithm involved in the computation of $S({\bm{c}})$ changes with the argument $\bm{c}$ and it is equal to the number of unique values in $\bm{c}$ so that $S({\bm{c}}) = 1$ can be obtained for any number of non-empty clusters $K_n\geq2$ (provided that $n/K_n\in \mathbb{N}$).
Clearly, when $\lambda$ is positive, for any candidate estimate $\hat c$, the loss function is inflated in correspondence of partitions $\bm{c}$ with high entropy, as desired.
  
Minimizing the expected entropy-regularized loss function $\bar L(\bm{c}, \hat{\bm{c}})$ with respect to the posterior is equivalent to minimizing the original loss function $L(\bm{c}, \hat{\bm{c}})$ with respect to an entropy-regularized version $\bar{\mathbb{P}}[\bm{c}\mid y_{1:n}] $ of the  posterior distribution, i.e.
\[
\bar{\mathbb{P}}[\bm{c}\mid y_{1:n}] \propto \exp\{\lambda\,S(\bm{c})\}\mathbb{P}[\bm{c}\mid y_{1:n}].
\]
This result, while immediate to prove, is highly desirable, because it allows implementation of the entropy-correction in a very straightforward and computationally feasible way which is described in Algorithm~\ref{alg}. 
Before computing summaries of the posterior, a resampling step is applied. 

More precisely, each sample from the posterior is resampled with probability proportional to $\exp\{\lambda\, S(\bm{c})\}$ so that an entropy-regularized version of the whole posterior distribution is obtained, thanks to a sampling importance resampling step.
Then, in the last step of the algorithm the original loss function $L(\bm{c}, \hat{\bm{c}})$ is minimized with respect to the entropy-regularized version of the posterior. 
Note that to solve such an optimization step we can rely on any of the effective algorithms available in the literature for the optimization of non-entropy regularized losses.
See e.g.\ \cite{rastelli2018optimal,dahl2022salso}.
In particular, we use the greedy algorithm described in \cite{dahl2022search} as implemented in the \texttt{R} library \textsf{salso} \citep{dahl2022salso} to perform the analysis presented in this work.

Thanks to the properties of the importance sampling procedure, the point estimate obtained minimizing $\sum_{m = 1 }^M L(\tilde {\bm{c}}_m, \hat{\bm{c}})$ is asymptotically equivalent to the solution of the minimization problem $\sum_{m = 1 }^M \bar L( {\bm{c}}_m, \hat{\bm{c}})$. However, there is a potential drawback of Algorithm~\ref{alg}, which stems from the finite dimension $M$ of the original sample from the posterior, ${\bm{c}_m, m=1,\ldots, M}$. Although Algorithm~\ref{alg} is easy to implement, it may significantly reduce the number of MCMC iterations considered in the minimization problem. To overcome this issue,  one possible solution may be to monitor the effective sample size (ESS) of the importance sampling step of Algorithm~\ref{alg} \citep[see e.g.,][]{liu1996metropolized} that can be approximated as
\begin{equation}\label{ESS}
\mbox{ESS} = \frac{1 }{ \sum_{m=1}^M  w_m^2}
\end{equation}
where $w_m$ are defined in Algorithm~\ref{alg} and $M$ is the number of initial draws for the posterior.
When the ESS is below a certain threshold, it can be increased by increasing the number of initial draws $M$ from the posterior.  
It is important to note that this use of the ESS indicator deviates from the conventional practice. 
The ESS is typically employed to measure the mixing performance of sampling algorithms, having as optimal value for the relative effective sample size $\text{ESS}/M$ the value of $1$. In this standard use, the ESS can be interpreted as the approximate number of independent draws obtained from a target distribution. However, this is not the case in our context. 
In fact, we should always expect a relative sample size $\text{ESS}/M$ lower than 1 to ensure that the entropy regularization has the desired effect on the estimates, the lower $ESS/M$ the higher the effect of the regularization. Moreover, here there is no target distribution we are referring to while computing the ESS. 
Roughly speaking, the entropy regularization shifts the \emph{importance} (i.e., the posterior density) towards specific areas in the support of the posterior and the ESS in \eqref{ESS} may serve only as a practical indicator of how well those areas have been previously explored by the original chain.

Finally, note that the choice of $\lambda$ plays an important role in defining the clustering estimator (as well as the choice of the not-regularized loss and the probabilistic clustering model assumptions).
The hyperparameter $\lambda$ can be elicited jointly with the rest of the prior settings, (e.g., prior, likelihood, and loss) in a Bayesian decision framework according to the preference on the point estimate of the clustering.
In particular, we recommend choosing $\lambda$ large if we want a stronger regularization.
How large depends on the specific analysis and the other model and prior choices.
In practice, if the goal is to use clustering just as a data summary can be easy and meaningful to try different values of $\lambda$ on a grid and see what produces more interpretable results in a cross-validation spirit.


\section{Simulation studies}

\label{sec:4}

\subsection{Univariate Gaussian mixtures}

\label{sec: gauss}

\begin{figure}[b!]
\begin{subfigure}[b]{0.32\textwidth}
	\centering
	\includegraphics[width=\textwidth]{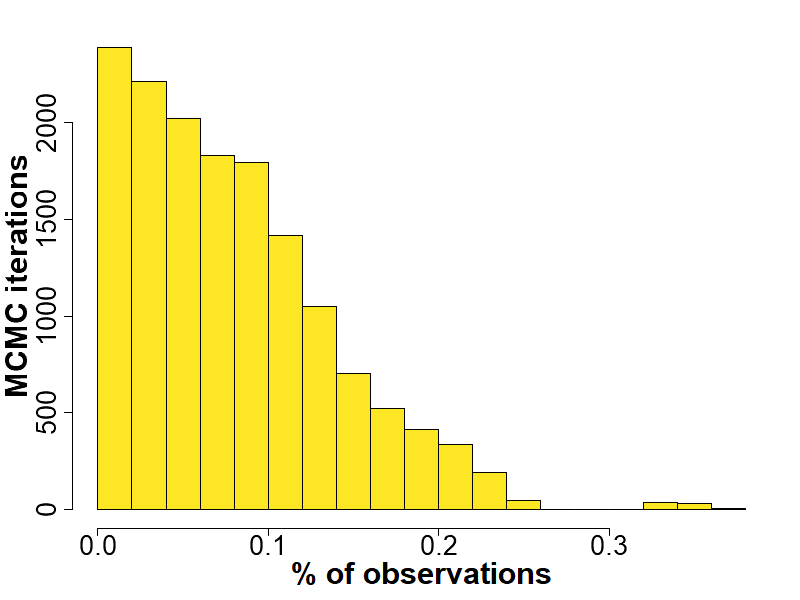} 
		\caption{Without entropy regularization. \label{fig:hist} }
\end{subfigure}
\hfill
	\begin{subfigure}[b]{0.32\textwidth}
	\centering\includegraphics[width=\textwidth]{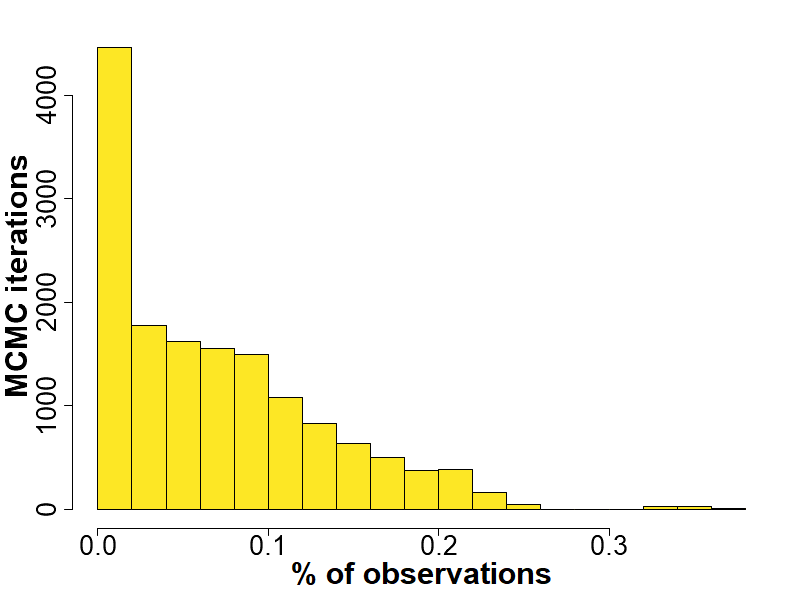}
	\caption{\label{fig:histreg10} With entropy regularization for $\lambda =10$.}
\end{subfigure}
\hfill
	\begin{subfigure}[b]{0.32\textwidth}
		\centering
    \includegraphics[width=\textwidth]{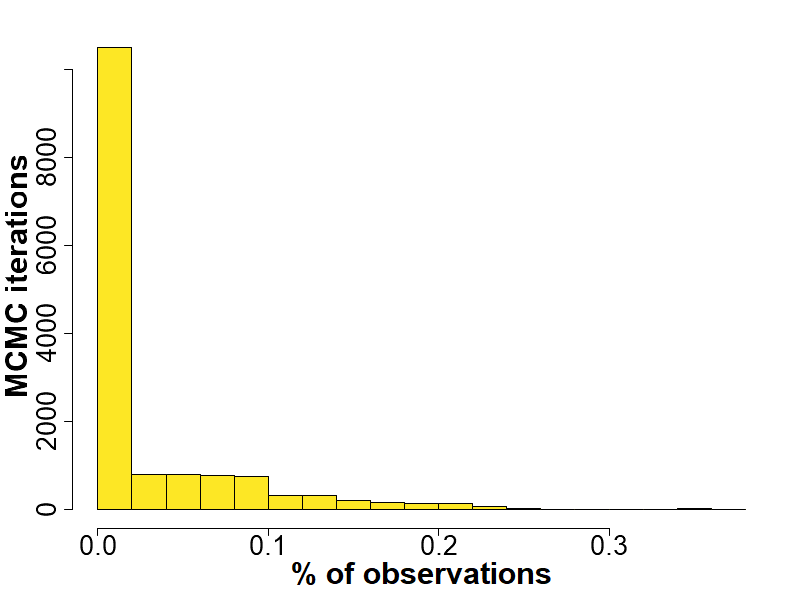}
		\caption{\label{fig:histreg} With entropy regularization for $\lambda =20$.}
\end{subfigure}
\caption{\label{fig:x}  
Gaussian simulation study.  
Percentage of observations in sparsely-populated clusters before and after entropy-regularization. 
Sparsely populated clusters are here defined as clusters containing 10\% or less of observations. 
The horizontal axis denotes the percentage of observations (out of 1000) that are assigned to those clusters. 
The y-axis represents the number of MCMC samples (out of 15 000).}
\end{figure}

We provide here a simulation study, where $n=1000$ observations are sampled from 3 distinct and well-separated univariate Gaussian distributions centered in $-4$, $0$, and $4$ and with unitary variance.
Here we refer to ``ground true'' clustering as the one implied by the membership indicators of the Gaussian kernels under the data-generating truth.
We employ a Normal-Normal DPM, with the base distribution centered at 0 and variance equal to 1. 
We compare the posterior estimates obtained by minimizing the Binder loss function and the entropy-regularized Binder loss function. 
We set the concentration parameter $\alpha = 1$, perform 20 000 MCMC simulations, and use the first 5000 as burn-in. 
See Section \ref{Appendix: A.2} for the results in the exact same setting but with a Gamma hyperprior for the concentration parameter.

Defining as sparsely populated clusters those clusters containing 10\% or less of observations, we found that in almost a third (4755 out 15 000) of the MCMC iterations, 10\% or more of the observations are allocated into sparsely populated clusters, while in almost two thirds (9306 out of 15 000) of MCMC iterations, 5\% or more of the observations are allocated into sparsely populated clusters, see Figure~\ref{fig:hist}. 
The same counts after entropy-regularization of the posterior (as described in the previous section) are, with $\lambda = 10$, 4088 and 7888 out 15 000, see Figure~\ref{fig:histreg10}, and, with $\lambda = 20$, 1375 and 3290 out 15 000, see Figure~\ref{fig:histreg}. 
Notice that coherently with the interpretation of the regularization in terms of the loss function, the regularized posterior should be intended only as a computational tool to provide summaries of the posterior distribution (e.g., point estimates and credible balls) and not as a posterior distribution itself.
\begin{figure}[tb]
\vspace{0.2cm}
	\begin{subfigure}[t]{0.24\textwidth}
		\centering
  	\includegraphics[width=\textwidth]{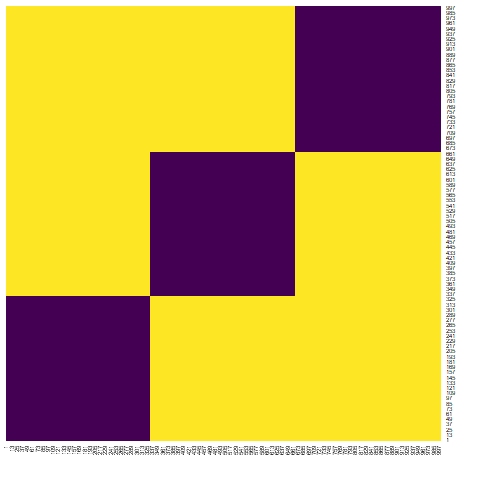}
		\caption{True clustering.}
	\end{subfigure}	
 \hfill
	\begin{subfigure}[t]{0.24\textwidth}
		\centering
  	\includegraphics[width=\textwidth]{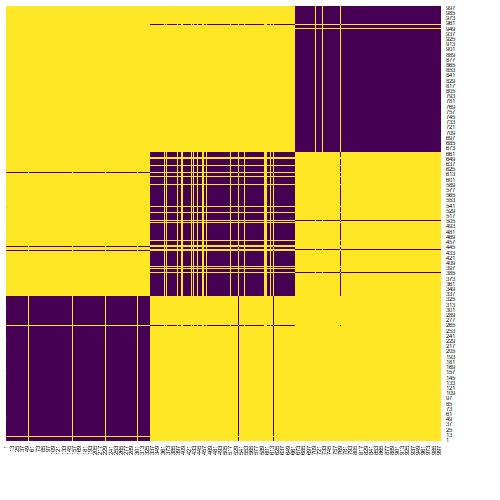}
		\caption{Binder loss clustering.}
	\end{subfigure}	
 \hfill
	\begin{subfigure}[t]{0.24\textwidth}
		\centering
	\includegraphics[width=\textwidth]{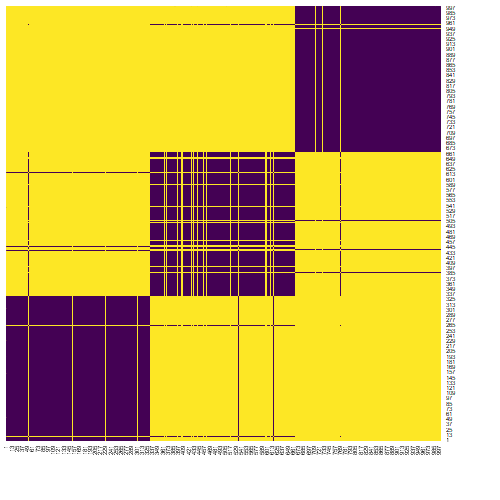}	
		\caption{Entropy regularized Binder loss clustering $\lambda = 10$.}
	\end{subfigure}	
 \hfill
\begin{subfigure}[t]{0.24\textwidth}
	\centering
 \includegraphics[width=\textwidth]{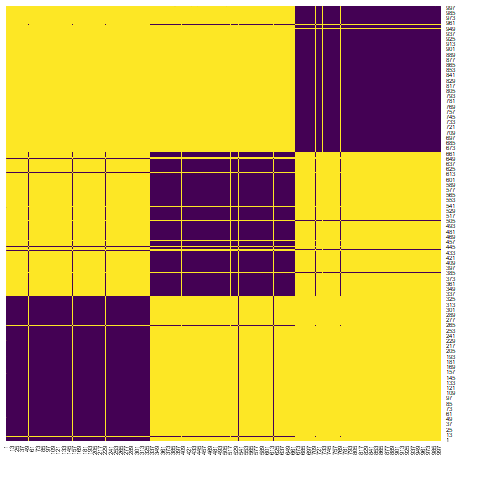}
	\caption{Entropy regularized Binder loss clustering $\lambda = 20$.}
\end{subfigure}	
	\caption{\label{fig:estsim} Gaussian simulation study.
    Estimated clustering for the simulation study darker squares denote couples of observations clustered together. 
    Panel (a) shows the ground true clustering. Panel (b) shows the clustering minimizing the Binder loss. 
    Panel (c) shows the clustering minimizing the entropy-regularized Binder loss for $\lambda = 10$ and panel (d) for $\lambda = 20$.}
\end{figure}

\begin{figure}[tb]
\vspace{0.2cm}
	\begin{subfigure}[b]{0.32\textwidth}
		\centering
  \includegraphics[width=\textwidth]{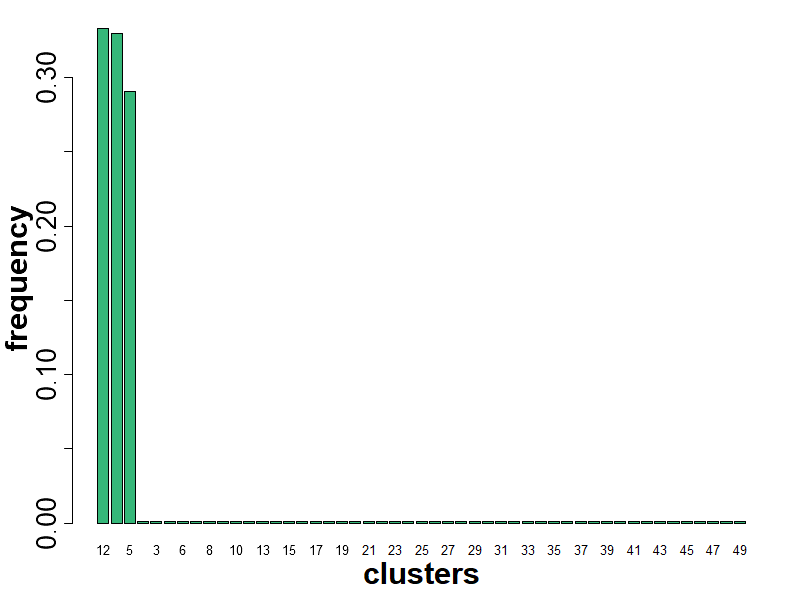}
		\caption{Without entropy regularization.}
	\end{subfigure}	
\hfill
	\begin{subfigure}[b]{0.32\textwidth}
		\centering
  \includegraphics[width=\textwidth]{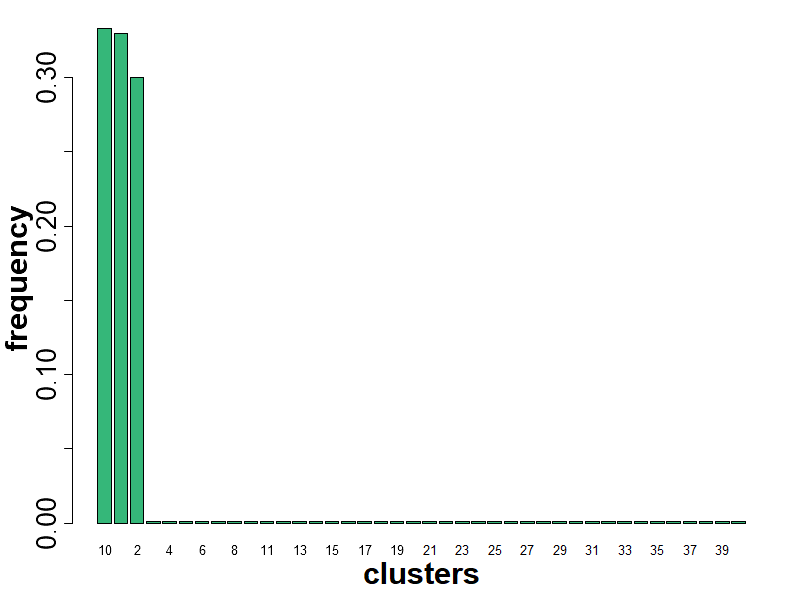}
		\caption{With entropy regularization for $\lambda = 10$.}
	\end{subfigure}	
\hfill
	\begin{subfigure}[b]{0.32\textwidth}
		\centering
  	\includegraphics[width=\textwidth]{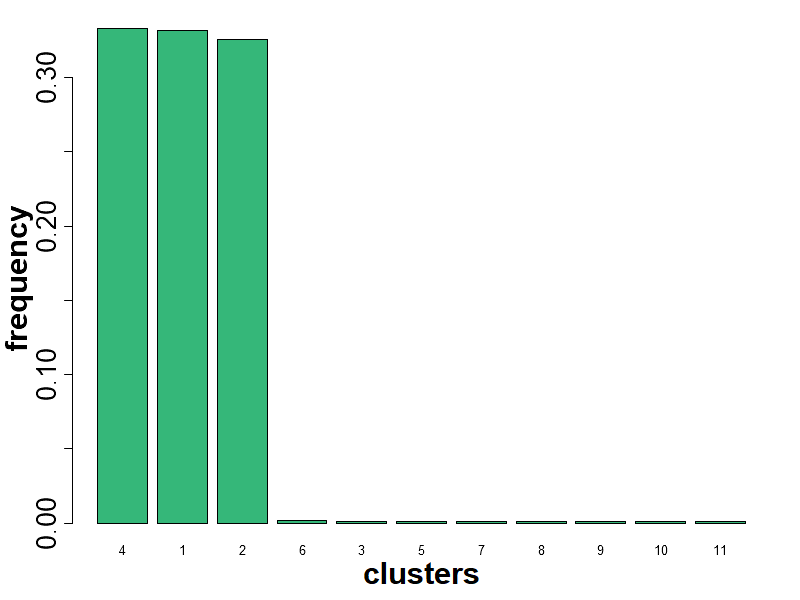}
		\caption{With entropy regularization for $\lambda = 20$.}
	\end{subfigure}	
	\caption{\label{fig:estfreq}  Gaussian simulation study.
 Estimated number of clusters and clusters' frequencies.}
\end{figure}

Finally, Figure~\ref{fig:estsim} shows the ground truth and the estimated clusters with and without entropy regularization. 
They highlight how the regularization allocates observations from noisy clusters into dominating ones. 
 
The main purpose of regularization is to provide a more interpretable and possibly more parsimonious representation of the dataset at hand without disregarding observations. 
Therefore, in general, the procedure prioritizes interpretability over the recovery of a frequentist truth.
Note also that even in the ideal frequentist situation of knowing the true data simulation density the misclassification rate will be typically low, but not zero, if the mixture kernels have overlapping supports as in the Gaussian scenario. 
However, monitoring misclassification errors in simulation studies can still be useful as it provides insights into how entropy regularization redistributes observations to achieve more balanced cluster frequencies. 
In this study, the application of regularization results in a reduction in misclassification errors, which is consistent with the fact that the highly unbalanced clusters are induced by the learning mechanism of BNP mixtures rather than the data itself. 
In particular, the number of observations misclassified (with respect to the simulation truth) with the Binder loss point estimate ($\lambda=0$) is $61$, with the regularization with $\lambda=10$ is $54$ and with $\lambda=20$ is $30$.

Figure~\ref{fig:estfreq} shows the cluster frequencies for the three point-estimates.
Note that in this simple univariate Gaussian kernel simulation scenario we can obtain the correct number of occupied components (i.e., clusters) and thus avoid sparsely populated clusters in the point estimate also using the VI loss (with the default parameter $a=1$ as implemented in \textsc{salso}) instead of considering the Binder loss (with the default parameter $a=1$).
However, without entropy regularization, both the VI loss and the Binder loss entail sparsely populated clusters in more complex scenarios, such as the multivariate simulation study presented in the next Section~\ref{sec: mult sim} and real-world dataset considered in the Section~\ref{sec:5}, respectively. 

\subsection{Multivariate Bernoulli mixtures and latent class analysis}
\label{sec: mult sim}

\begin{figure}[b!]
\begin{subfigure}[b]{0.32\textwidth}
	\centering
	\includegraphics[width=\textwidth]{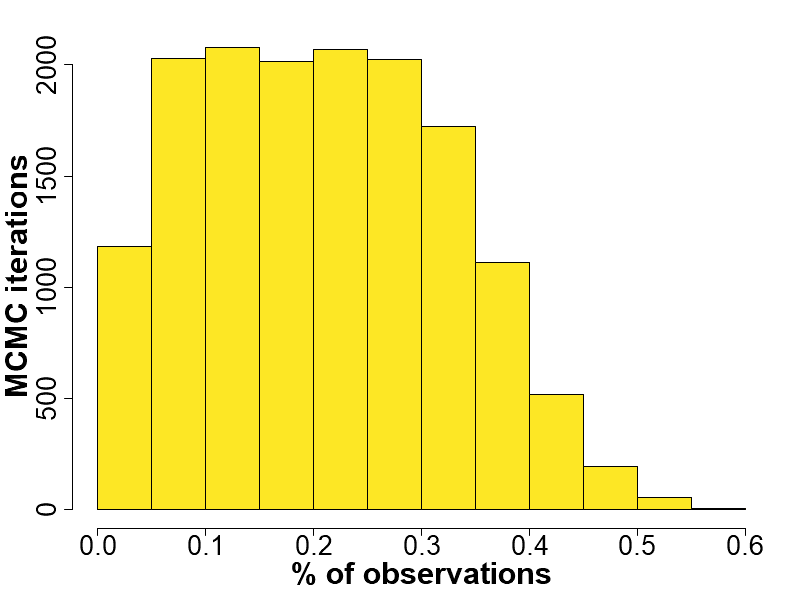} 
		\caption{Without entropy regularization. \label{fig:hist_bern} }
\end{subfigure}
\hfill
	\begin{subfigure}[b]{0.32\textwidth}
	\centering
\includegraphics[width=\textwidth]{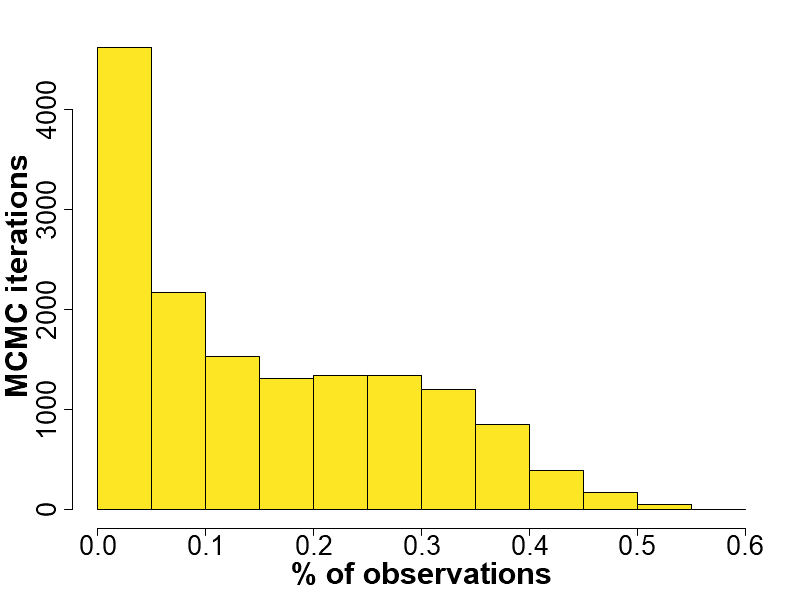}
	\caption{\label{fig:histreg10_bern} With entropy regularization for $\lambda =10$.}
\end{subfigure}
\hfill
	\begin{subfigure}[b]{0.32\textwidth}
		\centering
    \includegraphics[width=\textwidth]{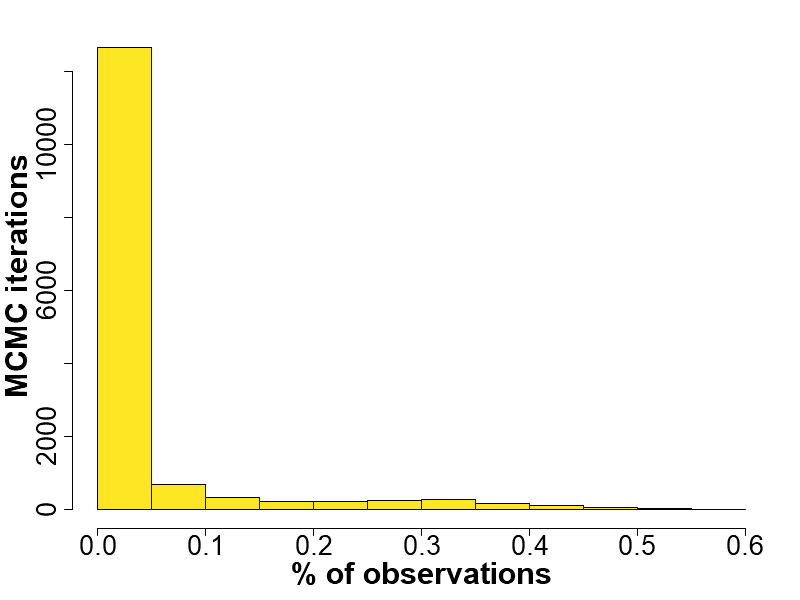}
		\caption{\label{fig:histreg_bern} With entropy regularization for $\lambda =20$.}
\end{subfigure}
\caption{\label{fig:x_bern}  
Multivariate Bernoulli simulation study.  
Percentage of observations in sparsely-populated clusters before and after entropy-regularization. 
Sparsely populated clusters are here defined as clusters containing 10\% or less of observations. 
The horizontal axis denotes the percentage of observations (out of 1000) that are assigned to those clusters. 
The y-axis represents the number of MCMC samples (out of 15 000).}
\end{figure}

In this section, we discuss the results of a synthetic numerical experiment involving multivariate Bernoulli data. Results are obtained employing a DPM with independent Bernoulli kernels, such that the likelihood is
\[
y_i\mid (w_h)_{h\geq1}, (p_{1,h},\ldots,p_{J,h})_{h\geq1} \overset{iid}{\sim} \sum_{h=1}^{\infty} w_h \left(\prod_{j=1}^J p_{j,h}^{y_i}(1-p_{j,h})^{1-y_i} \right)
\]
where the $h$th component has weight $w_h$ and it is characterized by the vector of probabilities 
$(p_{1,h},\ldots,p_{J,h})$. 
The model allows for the estimation of latent classes, which are commonly used in latent class analysis (LCA) to analyze multivariate discrete outcomes, often binary in nature.

\begin{figure}[b!]
\vspace{0.2cm}
	\begin{subfigure}[b]{0.32\textwidth}
		\centering
  \includegraphics[width=\textwidth]{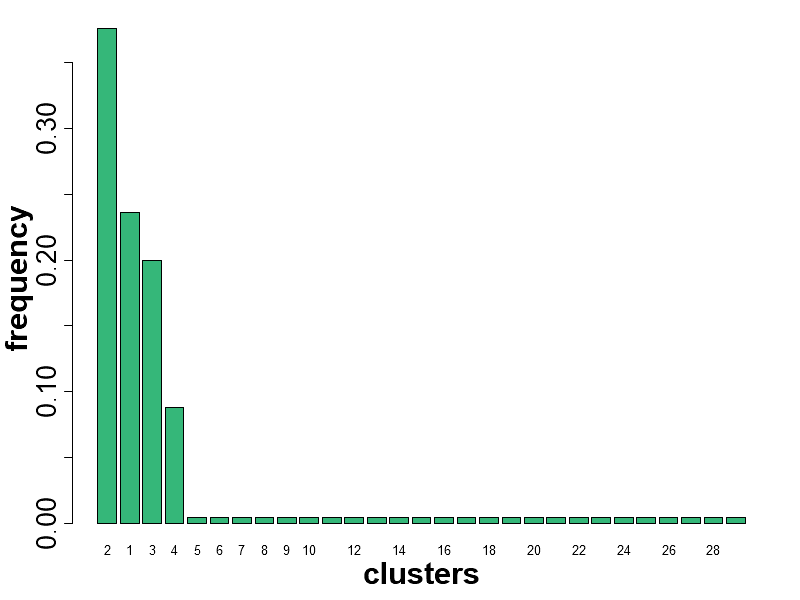}
		\caption{Without entropy regularization (Binder).}
	\end{subfigure}	
\hfill
	\begin{subfigure}[b]{0.32\textwidth}
		\centering
  \includegraphics[width=\textwidth]{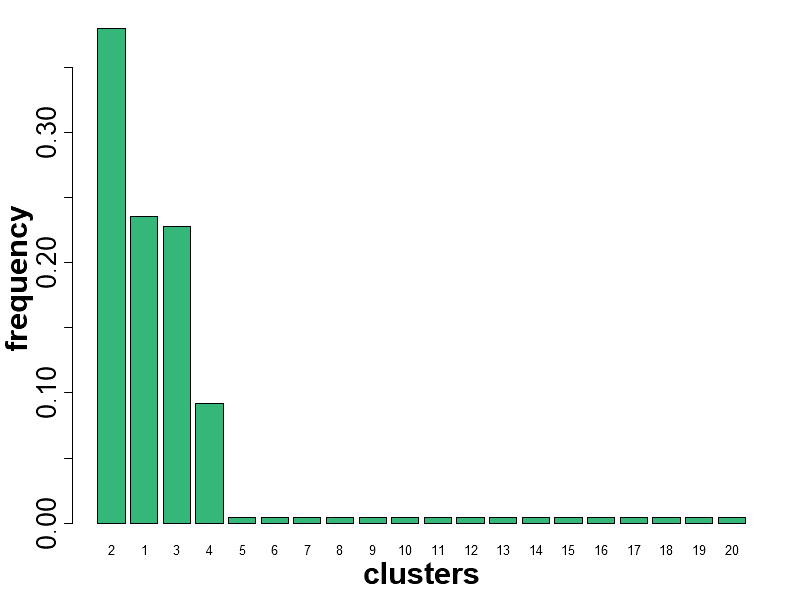}
		\caption{With entropy regularization for $\lambda = 10$ (Binder).}
	\end{subfigure}	
\hfill
	\begin{subfigure}[b]{0.32\textwidth}
		\centering
  	\includegraphics[width=\textwidth]{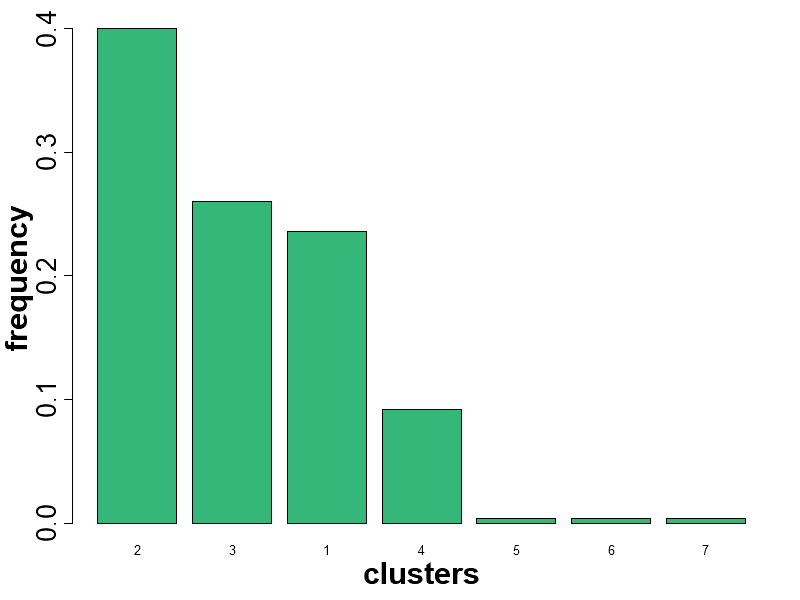}
		\caption{With entropy regularization for $\lambda = 20$ (Binder).}
	\end{subfigure}	
 \vspace{0.2cm}
	\begin{subfigure}[b]{0.32\textwidth}
		\centering
  \includegraphics[width=\textwidth]{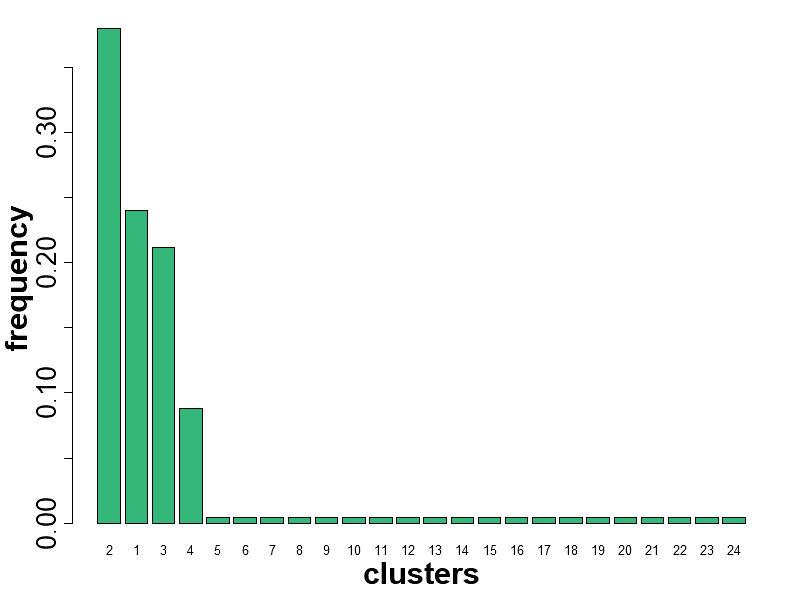}
		\caption{Without entropy regularization (VI).}
	\end{subfigure}	
\hfill
	\begin{subfigure}[b]{0.32\textwidth}
		\centering
  \includegraphics[width=\textwidth]{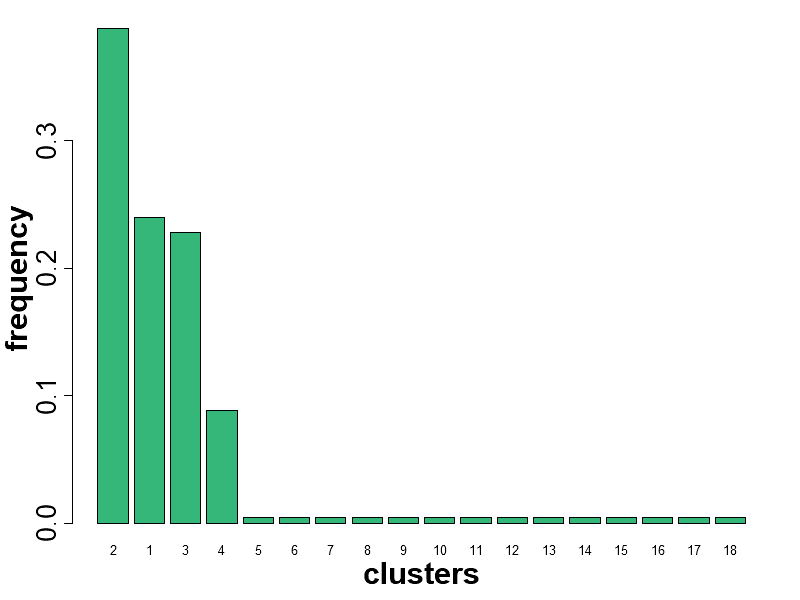}
		\caption{With entropy regularization for $\lambda = 10$ (VI).}
	\end{subfigure}	
\hfill
	\begin{subfigure}[b]{0.32\textwidth}
		\centering
  	\includegraphics[width=\textwidth]{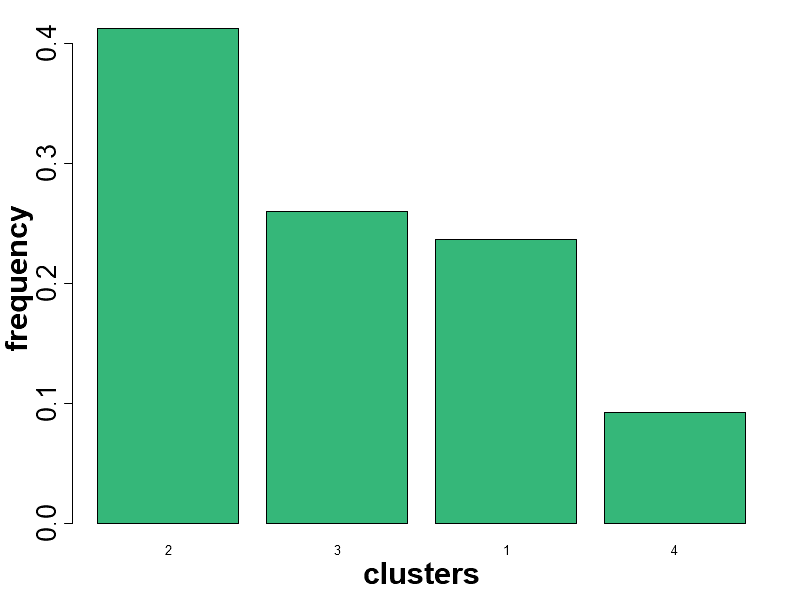}
		\caption{With entropy regularization for $\lambda = 20$ (VI).}
	\end{subfigure}	
	\caption{\label{fig:estfreq_bern}  Multivariate Bernoulli simulation study.
 Estimated number of clusters and clusters' frequencies.}
\end{figure}

In LCA, each latent class is represented by a mixture component, and the observations within each class are assumed to be independent. 
This assumption holds for LCA even though, typically, the observed variables are assumed to be statistically dependent. 
This is a crucial aspect of LCA: the classes are indeed used to represent the observed dependence \citep[see, for instance,][]{mccutcheon1987latent}. 
The rationale behind the approach is that estimating the dependence across $J$ binary outcomes is often challenging, particularly when $J$ is large, as a $J$-variate Bernoulli distribution requires $J^2-1$ parameters to be estimated. 
The goal of LCA is to explain and approximate the observed dependence in the data by introducing latent classes.
Thus, this model serves the purpose of approximating complex dependent $J$-variate binary distributions through the identification of patterns in the data that can explain the observed dependence in a more concise and interpretable manner than estimating the entire set of $J^2-1$ parameters. 
This method is also referred to as ``the categorical data analog of factor analysis" \citep{mccutcheon1987latent}. 
For more details on classical LCA we refer to \cite{lazarsfeld1955recent}, \cite{mccutcheon1987latent}, and \cite{andersen1982latent}, for Bayesian LDA to \cite{white2014bayeslca} and \cite{li2018bayesian}, and for recent Bayesian nonparametric generalizations to \cite{bartolucci2017nonparametric}, \cite{koo2020bayesian},  \cite{franzolini2023bayesian}, and \cite{qiu2023bayesian}.

In this simulation, we generate data for $n=250$ subjects and $p=50$ positively correlated binary outcomes, with pairwise correlations ranging from $0.0871$ to $0.5014$. 
We fit a multivariate Bernoulli DPM, with a J-variate product of $\text{Beta}\,(0.2, 0.2)$ as base distribution and a $\text{Gamma}\,(1,1)$ prior on the concentration parameter $\alpha$, perform $20 \, 000$ MCMC simulations, and use the first $5 \,000$ as burn-in.

Defining, as in the previous section, as sparsely populated clusters those clusters containing 10\% or less of observations, we have that in $11\,788$ out $15\,000$ of the MCMC iterations, $10\%$ or more of the observations are allocated into sparsely populated clusters, while in $13\,815$ out of $15\,000$ of MCMC iterations, $5\%$ or more of the observations are allocated into sparsely populated clusters, see Figure~\ref{fig:hist_bern}. 
The same counts after entropy-regularization of the posterior are, with $\lambda = 10$, $8\,203$ and $10\,373$ out $15\,000$, see Figure~\ref{fig:histreg10_bern}, and, with $\lambda = 20$, $1\,660$ and $2\,342$ out $15\,000$, see Figure~\ref{fig:histreg_bern}. Figure~\ref{fig:estfreq_bern} shows the cluster frequencies for the three point-estimates obtained with both Binder and VI losses, without and with regularization.

\section{Results for the wine dataset}
\label{sec:5}
\begin{figure}[ht]
    \centering
	\begin{subfigure}[h]{0.3\textwidth}
		\centering
  	\includegraphics[width=\textwidth]{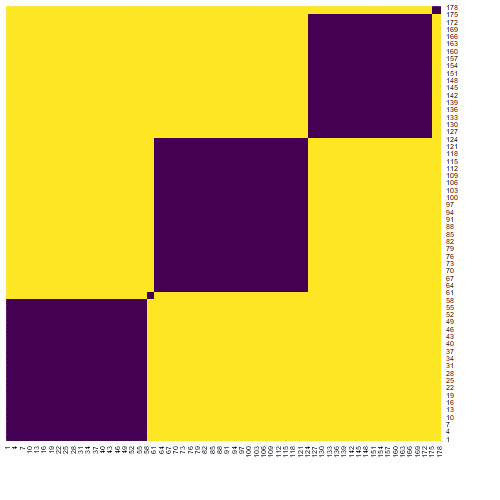}
	\caption{Estimated partition (Binder) without entropy regularization.}
	\end{subfigure}	
\hfill
\begin{subfigure}[h]{0.3\textwidth}
	\centering
    \includegraphics[width=\textwidth]{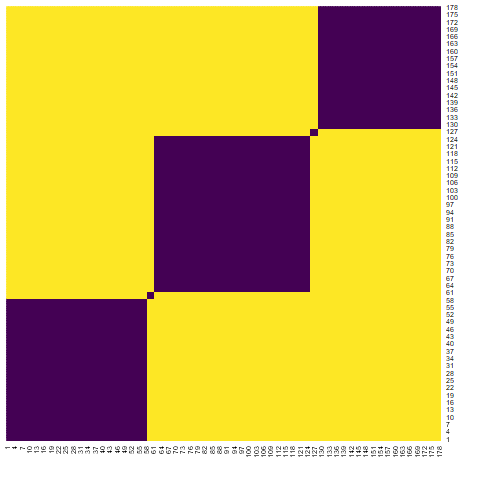}
    \caption{Estimated partition (VI) without entropy regularization.}
\end{subfigure}	
\hfill
	\begin{subfigure}[h]{0.3\textwidth}
	\centering
	\includegraphics[width=\textwidth]{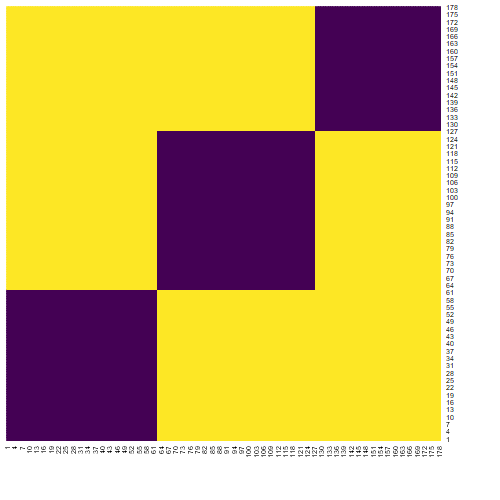}
	\caption{Estimated partition (same with Binder or VI) after entropy regularization.}
	\end{subfigure}	
\caption{\label{fig:estp} 
Estimated partitions for the wine dataset. 
Darker squares denote couples of observations clustered together, observations are ordered based on co-clustering. 
}
\end{figure}

We test the performance of our method also on the wine dataset available on \textsf{R}, where data are the results of a chemical analysis of wines grown in the same region in Italy but derived from three different cultivars. 
The analysis determined the quantities of 13 constituents found in each of the three types of wines. 
Here we refer to the clustering identified by the three types of wines as ``ground truth''.

\begin{figure}[!tb]
	\begin{subfigure}{0.3\textwidth}
		\includegraphics[width=\textwidth]{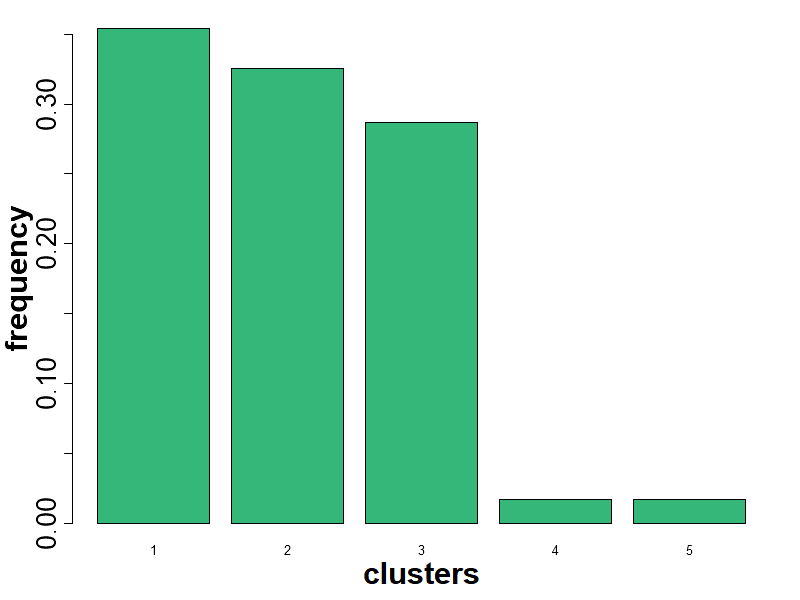}	
		\caption{Without entropy regularization (Binder).}
	\end{subfigure}	
\hfill
	\begin{subfigure}{0.3\textwidth}
		\includegraphics[width=\textwidth]{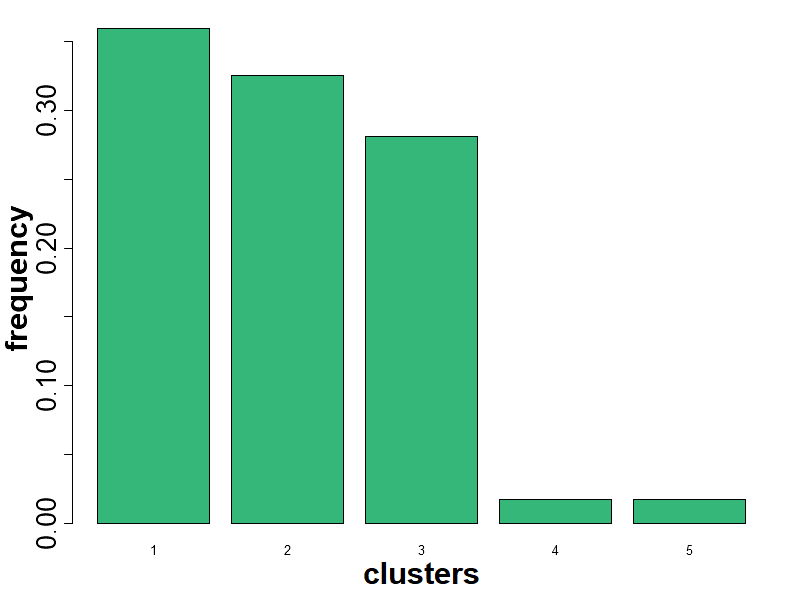}	
		\caption{Without entropy regularization (VI).}
	\end{subfigure}	
\hfill
	\begin{subfigure}{0.3\textwidth}
  	\includegraphics[width=\textwidth]{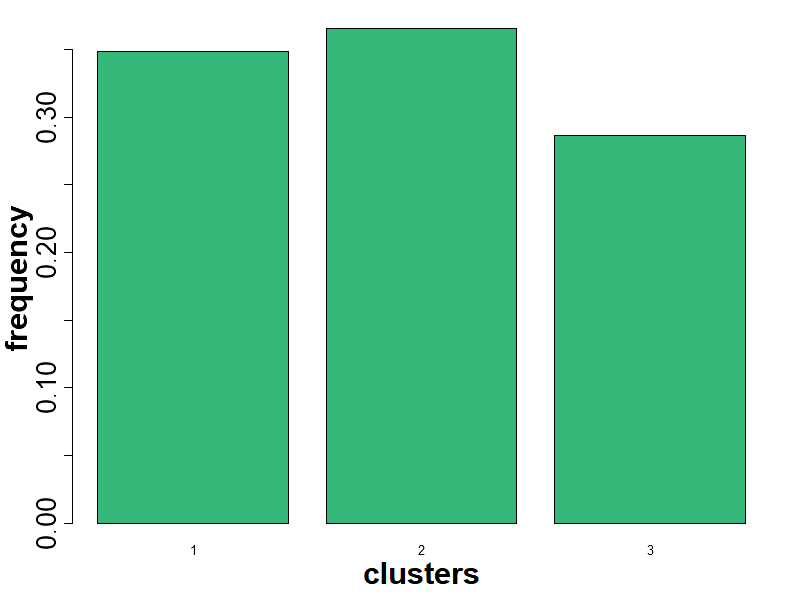}
		\caption{With entropy regularization (Binder or VI).}
	\end{subfigure}	
\caption{\label{fig:clustfreq} 
Estimated number of clusters and clusters' frequencies for the wine dataset.}
\end{figure}

We use the 13 constituents to estimate a Dirichlet process mixture model with a multivariate Gaussian kernel, and we try to recover the three groups of types of wine through the estimated clustering. 
The concentration parameter is set to $0.1$ to further favor a small number of clusters (see Section \ref{Appendix: A.1} for the results in the exact same setting but with a Gamma hyperprior for the concentration parameter). 
Data have been scaled before estimating the clustering configuration. 
After running the MCMC for 20 000 iterations and using the first 5000 as burnin, both the Binder loss and the VI functions identify a partition of five clusters, while our estimator for $\lambda = 50$ correctly identifies three clusters. 
Note that both the point estimates obtained with regularizing (with $\lambda=50$) the Binder loss and the VI loss are identical in this analysis (contrary to their not regularized estimates).
See Figure~\ref{fig:estp} and Figure~\ref{fig:clustfreq}. 

\begin{figure}[!tb]
	\begin{subfigure}[t]{0.24\textwidth}
		\centering
 \includegraphics[width=\textwidth]{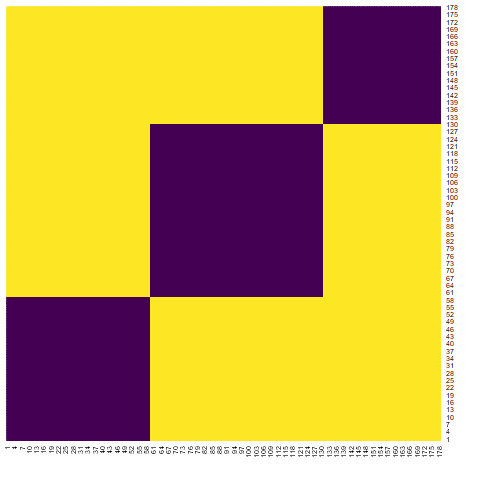}
		\caption{Types of wines.}
	\end{subfigure}	
 \hfill
	\begin{subfigure}[t]{0.24\textwidth}
		\centering
  \includegraphics[width=\textwidth]{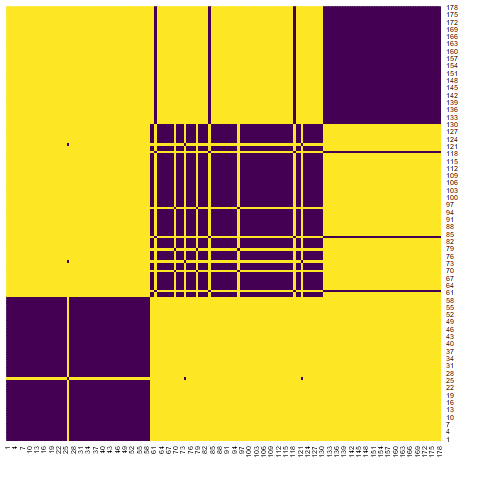}
		\caption{Binder loss.}
	\end{subfigure}	
  \hfill
	\begin{subfigure}[t]{0.24\textwidth}
		\centering
  \includegraphics[width=\textwidth]{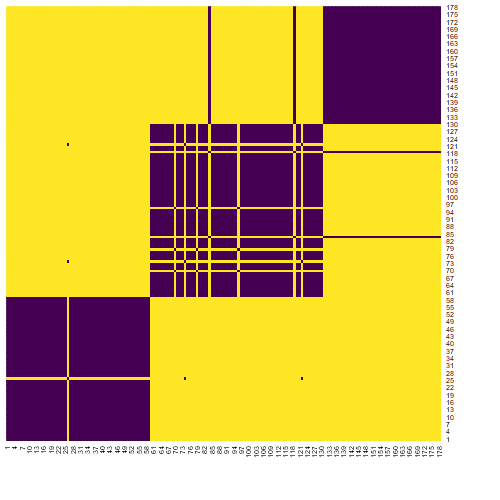}
		\caption{VI loss.}
	\end{subfigure}	
  \hfill
	\begin{subfigure}[t]{0.24\textwidth}
		\centering
  	\includegraphics[width=\textwidth]{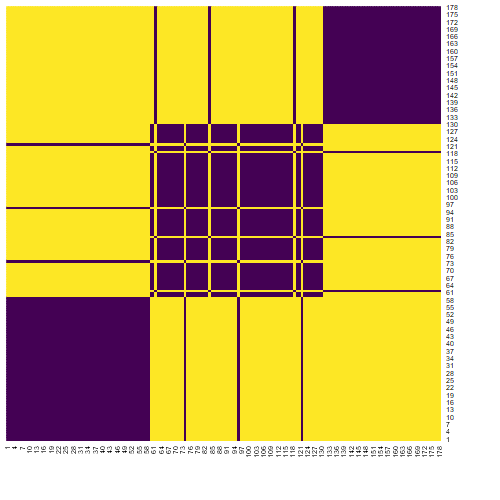}
		\caption{Entropy regularized (Binder or VI).}
	\end{subfigure}	
	\caption{\label{fig:4} 
 Estimated clustering for the wine dataset. 
 Darker squares denote couples of observations clustered together, observations are ordered based on three groups of types of wine. 
 }
\end{figure}

Lastly, Figure \ref{fig:4} compares the clustering based on three groups of types of wine with the three estimates.
The number of wrongly allocated wines, which equals 9 in the Binder loss point estimate and 8 in the VI point estimates, is reduced to 6 in the entropy regularized point estimate.

\FloatBarrier
\section{Conclusions}
As highlighted in the recent literature, common posterior point estimates of the clustering obtained from Bayesian discrete mixture models suffer from unbalanced clusters' frequencies with only a few dominating clusters and a large number of sparsely-populated ones.
In contrast, we introduced a general entropy-regularization of the existing losses that reduces the number of sparsely-populated clusters and enhances the interpretability of the Bayesian point estimate.
Importantly, our proposal is theoretically justified and does not break the projectivity of the Bayesian model. 
We have further devised a simple and general computational scheme allowing for efficient computation of such entropy-regularized clustering estimate.
This work paves the way for future intriguing research directions that we plan to address in forthcoming works.

From a theoretical perspective, it is interesting to study the connection with a wise recent probabilistic clustering model that induces less unbalanced clusters via breaking Kolmogorov consistency \citep{lee2022rich}.
Indeed we note that the entropy penalization introduced in this work can be incorporated in the prior  instead of being applied to the loss function, ultimately constituting a sparsity penalized random partition model.
However, such a random partition model breaks Kolmogorov consistency, contrary to our coherent loss-based approach discussed in Section \ref{sec:3}. 
The decision to use a Kolmogorov consistent or inconsistent model ultimately depends on the specific applied problem. 
One of the strengths of our loss-based approach lies in the fact that, regardless of the choice made in this regard by the analyst, the Kolmogorov consistency of the model would not be affected by adopting our technique. 
Nonetheless, this duality of our proposal allows bridging connection with other recent interesting non-Kolmogorov consistent random partition models that are built by modifying existing EPPFs.
See, e.g., \cite{dahl2017random, paganin2021centered, zanella2016flexible}.

From modeling and applied perspectives, it is natural to move beyond the exchangeable case and extend our regularized loss-based estimator to perform probabilistic clustering for dependent random partition models that allow considering covariates \citep[see e.g.,][]{teh2006hierarchical, muller2011product, page2022clustering} or time-dependent random partition models such as those proposed, for example, in \cite{page2022dependent, franzolini2023conditional}. 
Finally, the general loss-penalization that we have introduced and Algorithm~\ref{alg} seem an appropriate tool for performing joint probabilistic clustering of different entities like in separate exchangeable partition models that allow performing bi-clustering in matrix data \citep{lee2013nonparametric,lin2023separate} and nested random partition model that allows us to jointly cluster populations and observations such as in nested partial exchangeable partition models \citep{rodriguez2008nested,zuanetti2018clustering} and recent extensions. 

\section*{Acknowledgement}
The authors are grateful to the Editor and two
anonymous referees for insightful comments and suggestions.
B.\ Franzolini is supported by PNRR -
PE1 FAIR - CUP B43C22000800006.

\setcounter{equation}{0}
\setcounter{table}{0}
\setcounter{figure}{0}
\setcounter{section}{0}
\numberwithin{table}{section}
\renewcommand{\theequation}{A.\arabic{equation}}
\renewcommand{\thesection}{A.\arabic{section}}
\renewcommand{\thesubsection}{A.\arabic{section}.\arabic{subsection}}
\renewcommand{\thetable}{A.\arabic{table}}
\renewcommand{\thefigure}{A.\arabic{figure}}

\section{Appendix: Prior on the concentration parameter}
In Section \ref{sec:2} we report the partition cost associated with a DPM.
It is clear that the choice of the concentration parameter $\alpha$ is relevant in controlling the number of clusters.
Thus, in order to have a more flexible distribution on the clustering of the data, in many implementations of the Dirichlet process mixture 
a prior for $\alpha$ is specified, leading to a mixing measure that is itself a mixture in the sense of \cite{antoniak1974mixtures}. 
\cite{ascolani2023clustering} also show that introducing such a prior can have a major impact on the asymptotic behavior of the number of clusters, as Dirichlet process mixtures can be consistent for the number of clusters. 
In these sections, we show the results obtained by repeating the analyses performed in the main paper where we use a Gamma prior \citep{escobar1995bayesian} on the concentration parameter of the Dirichlet process.
More precisely, we change the prior such that
\begin{equation*}
    \alpha \sim \text{Gamma}(1,1)
\end{equation*}
and all the remaining prior specifications and MCMC settings (e.g., also the number of iterations) are set equal to the previous one.
We show that also in such a case the proposed entropy-regularization is still crucial to enhance the interpretability of the clustering point estimate.

\FloatBarrier
\subsection{Wine dataset}
\label{Appendix: A.1}
\begin{figure}[H]
    \centering
	\begin{subfigure}[h]{0.24\textwidth}
		\centering
  	\includegraphics[width=\textwidth]{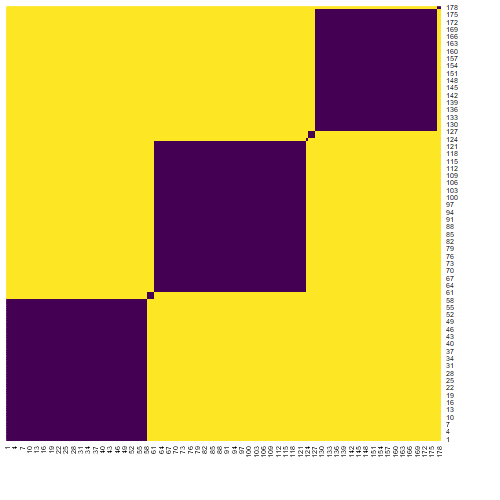}
	\caption{Estimated partition (Binder) without entropy regularization.}
	\end{subfigure}	
\hfill
\begin{subfigure}[h]{0.24\textwidth}
	\centering
    \includegraphics[width=\textwidth]{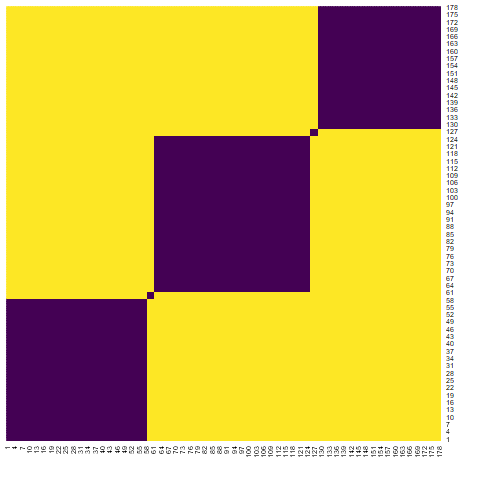}
    \caption{Estimated partition (VI) without entropy regularization.}
\end{subfigure}	
\hfill
	\begin{subfigure}[h]{0.24\textwidth}
	\centering
	\includegraphics[width=\textwidth]{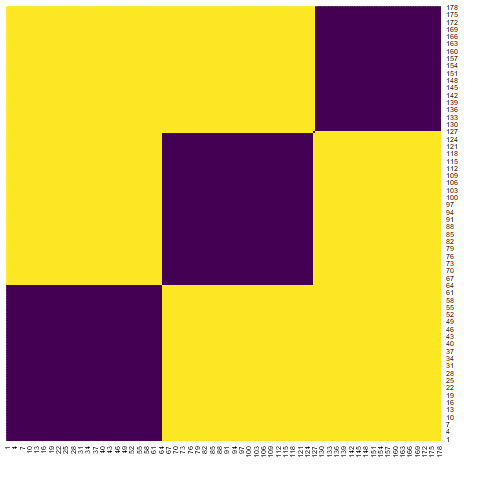}
	\caption{Estimated partition (Binder) after entropy regularization.}
	\end{subfigure}	
\hfill
 \begin{subfigure}[h]{0.24\textwidth}
	\centering
	\includegraphics[width=\textwidth]{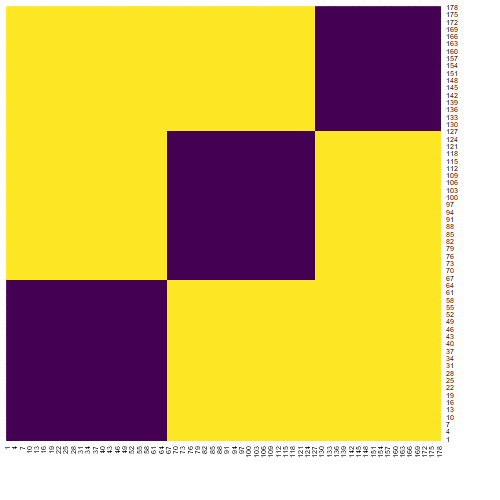}
	\caption{Estimated partition (VI) after entropy regularization.}
	\end{subfigure}	
\caption{\label{fig:estp_app} 
Estimated ($\alpha$ random) partitions for the wine dataset. 
Darker squares denote couples of observations clustered together, observations are ordered based on co-clustering. 
}
\end{figure}

\begin{figure}[H]
	\begin{subfigure}{0.24\textwidth}
		\includegraphics[width=\textwidth]{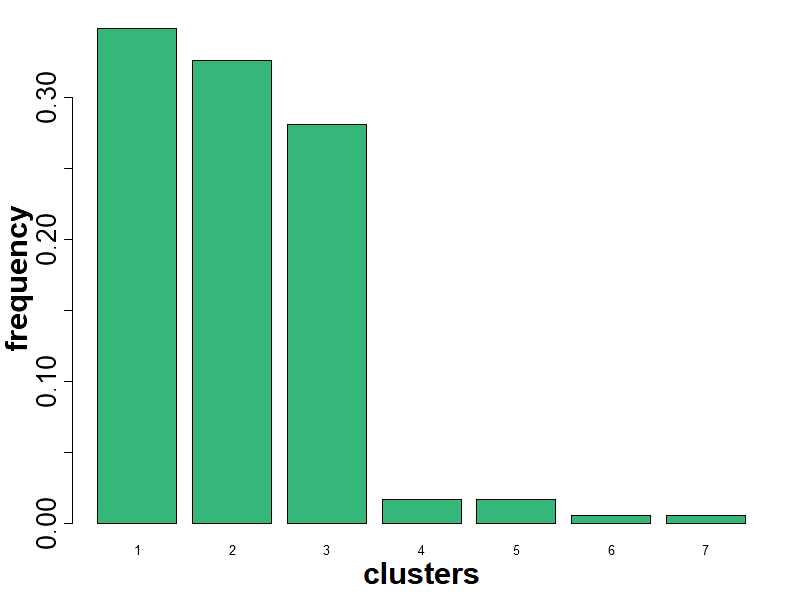}	
		\caption{\mbox{Without entropy} \mbox{regularization (Binder).}}
	\end{subfigure}	
\hfill
	\begin{subfigure}{0.24\textwidth}
		\includegraphics[width=\textwidth]{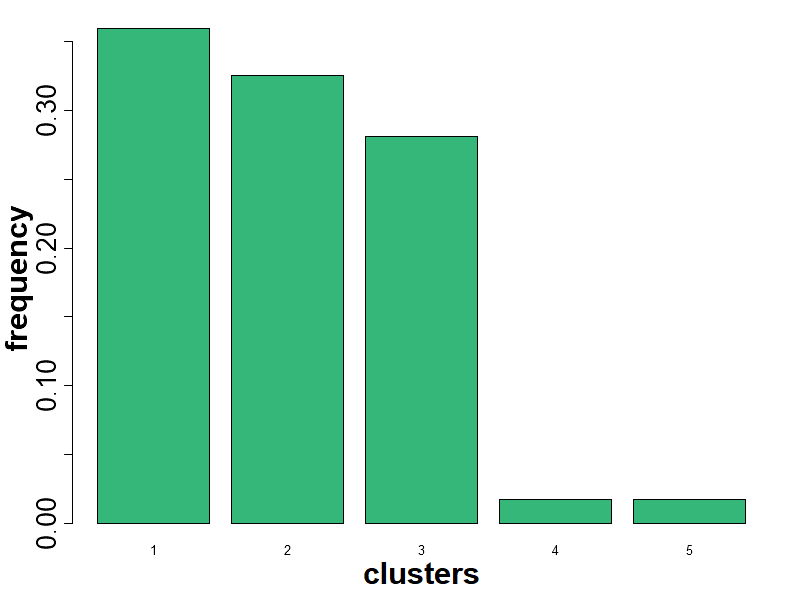}	
		\caption{Without entropy regularization (VI).}
	\end{subfigure}	
\hfill
	\begin{subfigure}{0.24\textwidth}
  	\includegraphics[width=\textwidth]{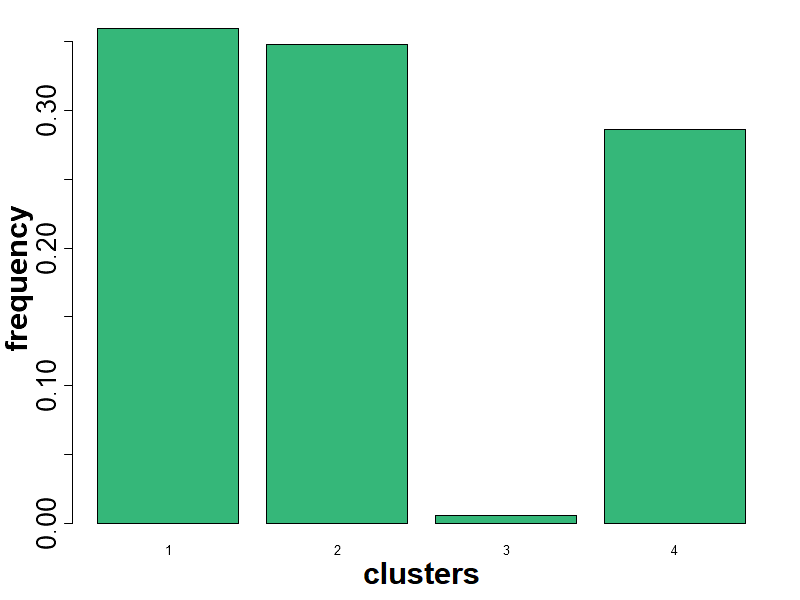}
		\caption{With entropy regularization (Binder).}
	\end{subfigure}	
 \hfill
	\begin{subfigure}{0.24\textwidth}
  	\includegraphics[width=\textwidth]{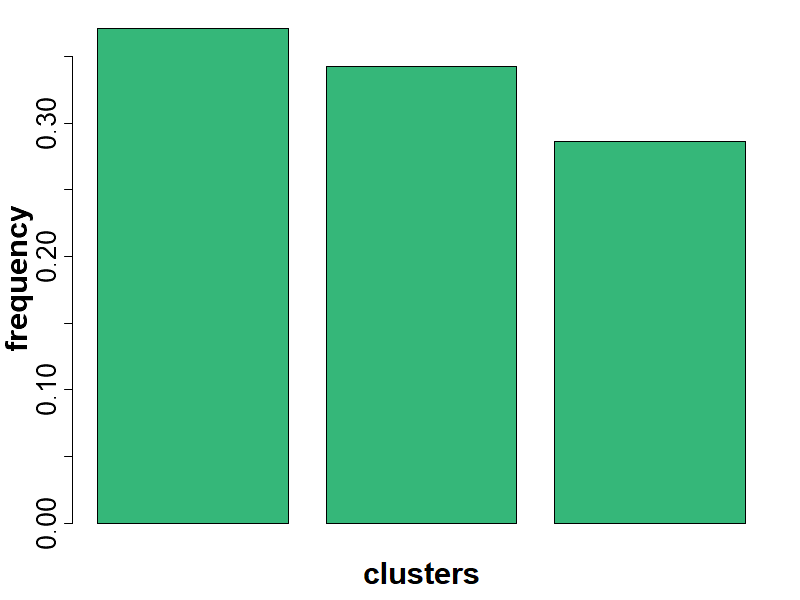}
		\caption{With entropy regularization (VI).}
	\end{subfigure}	
\caption{\label{fig:clustfreq_app} 
Estimated ($\alpha$ random) number of clusters and clusters' frequencies for the wine dataset.}
\end{figure}

\begin{figure}[H]
	\begin{subfigure}[t]{0.19\textwidth}
		\centering
 \includegraphics[width=\textwidth]{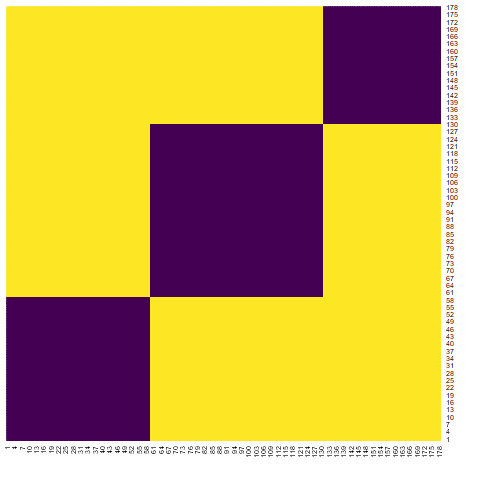}
		\caption{Types of wines.}
	\end{subfigure}	
 \hfill
	\begin{subfigure}[t]{0.19\textwidth}
		\centering
  \includegraphics[width=\textwidth]{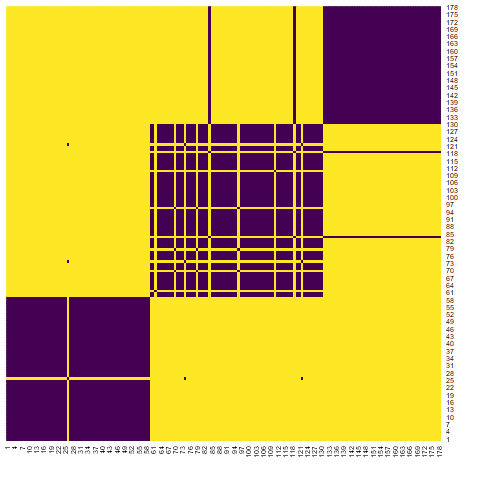}
		\caption{Binder loss clustering.}
	\end{subfigure}	
 \hfill
	\begin{subfigure}[t]{0.19\textwidth}
		\centering
  \includegraphics[width=\textwidth]{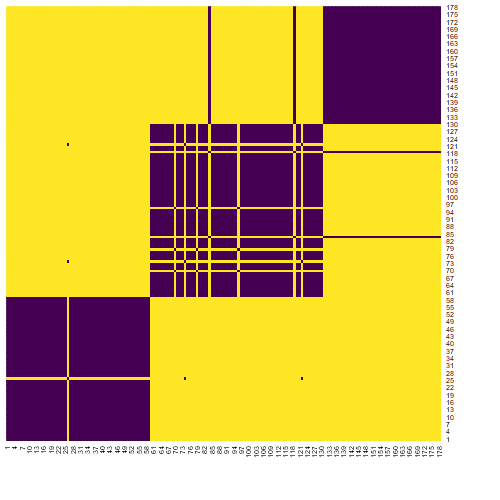}
		\caption{VI loss clustering.}
	\end{subfigure}	
 \hfill
	\begin{subfigure}[t]{0.19\textwidth}
		\centering
  	\includegraphics[width=\textwidth]{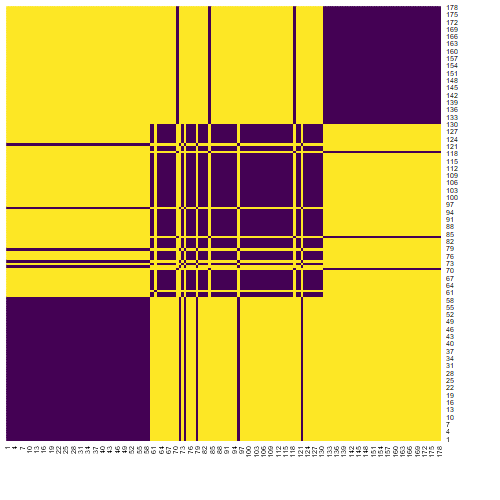}
		\caption{Entropy regularized clustering (Binder).}
	\end{subfigure}	
 \hfill
	\begin{subfigure}[t]{0.19\textwidth}
		\centering
  	\includegraphics[width=\textwidth]{Image/corrected_VI_wine_reorder_app.png}
		\caption{Entropy regularized clustering (VI).}
	\end{subfigure}	
	\caption{\label{fig:4_app} 
 Estimated clustering for the wine dataset ($\alpha$ random). 
 Darker squares denote couples of observations clustered together, observations are ordered based on three groups of types of wine. 
}
\end{figure}

We note that the point estimates obtained with regularizing (with $\lambda=50$) the Binder loss and the VI loss are different in this analysis (contrary to the not regularized estimates with fixed concentration parameter).
See Figure~\ref{fig:estp_app} and Figure~\ref{fig:clustfreq_app} for the estimated partitions and related frequencies, respectively. 
Figure \ref{fig:4_app} compares the clustering based on three groups of types of wine with the four estimates.
The number of wrongly allocated wines is equal to $10$ in the Binder loss point estimate and $8$ in the VI point estimates, while $8$ and $10$ with their regularized versions.
Moreover, the number of clusters in the point estimate of the partition is $7$ with the Binder loss and $5$ with the VI loss, while $4$ and $3$ with their regularized versions (where the ground-truth is 3).

\subsection{Gaussian simulation scenario}
\label{Appendix: A.2}
\begin{figure}[H]
\begin{subfigure}[b]{0.3\textwidth}
	\centering
	\includegraphics[width=\textwidth]{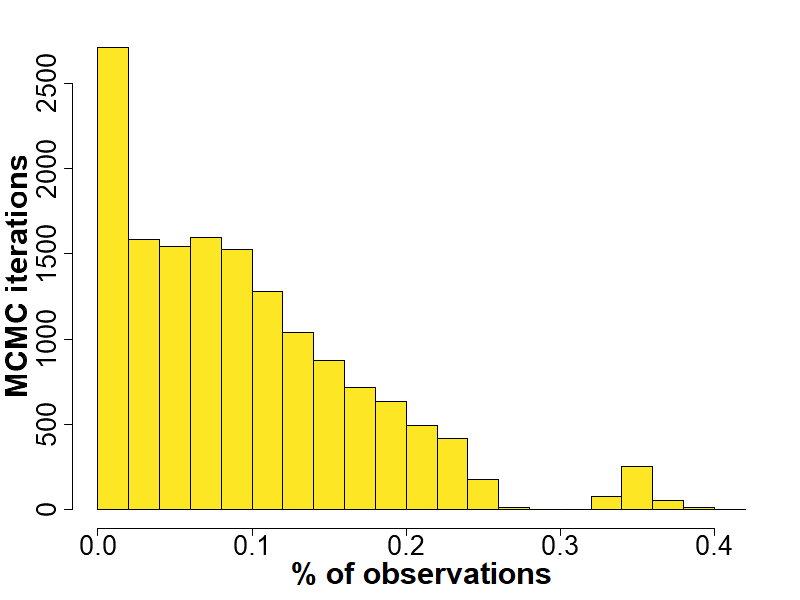} 
		\caption{Without entropy regularization. \label{fig:hist_app} }
\end{subfigure}
\hfill
	\begin{subfigure}[b]{0.3\textwidth}
	\centering
\includegraphics[width=\textwidth]{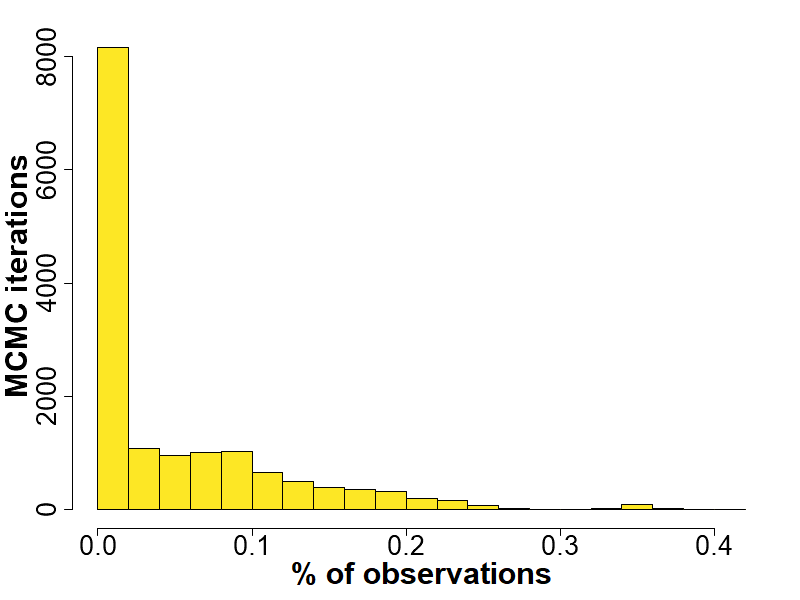}
	\caption{\label{fig:histreg10_app} With entropy regularization for $\lambda =10$.}
\end{subfigure}
\hfill
	\begin{subfigure}[b]{0.3\textwidth}
		\centering
    \includegraphics[width=\textwidth]{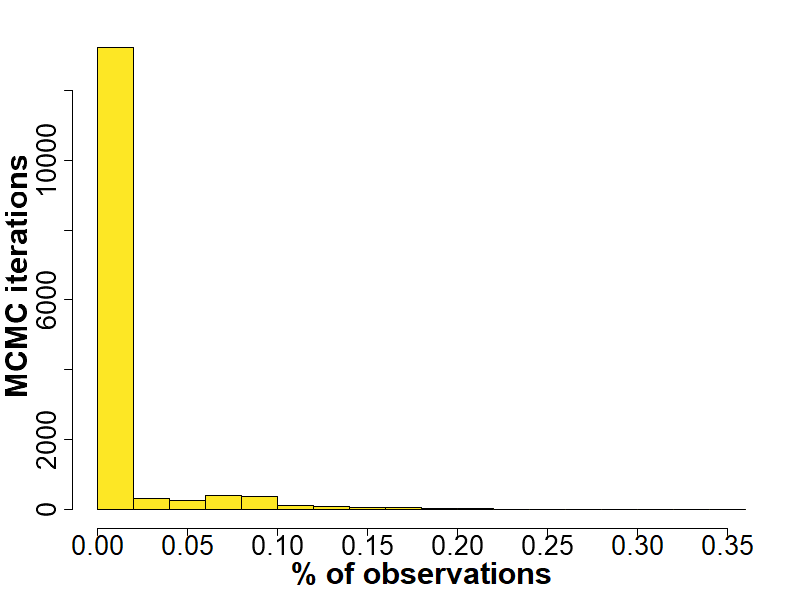}
		\caption{\label{fig:histreg_app} With entropy regularization for $\lambda =20$.}
\end{subfigure}
\caption{\label{fig:x_app}  
Gaussian simulation study ($\alpha$ random).  
Percentage of observations in sparsely-populated clusters before and after entropy-regularization. 
Sparsely populated clusters are here defined as clusters containing 10\% or less of observations.}
\end{figure}

\begin{figure}[H]
\vspace{0.2cm}
	\begin{subfigure}[t]{0.2\textwidth}
		\centering
  	\includegraphics[width=\textwidth]{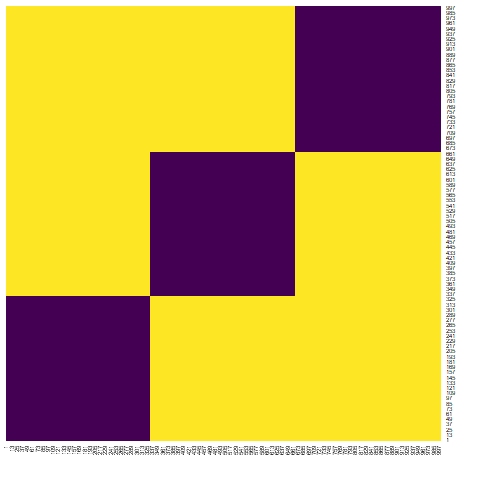}
		\caption{True clustering.}
	\end{subfigure}	
 \hfill
	\begin{subfigure}[t]{0.2\textwidth}
		\centering
  	\includegraphics[width=\textwidth]{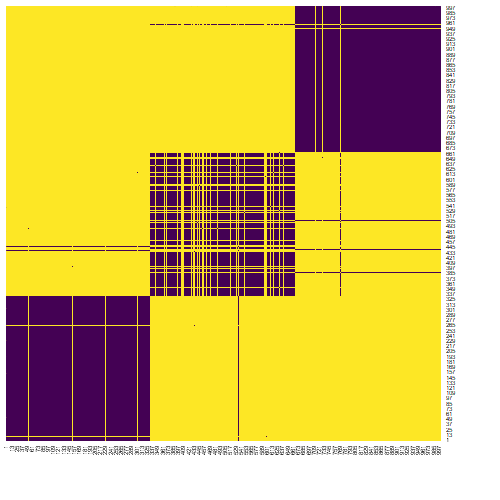}
		\caption{Binder loss clustering.}
	\end{subfigure}	
 \hfill
	\begin{subfigure}[t]{0.2\textwidth}
		\centering
	\includegraphics[width=\textwidth]{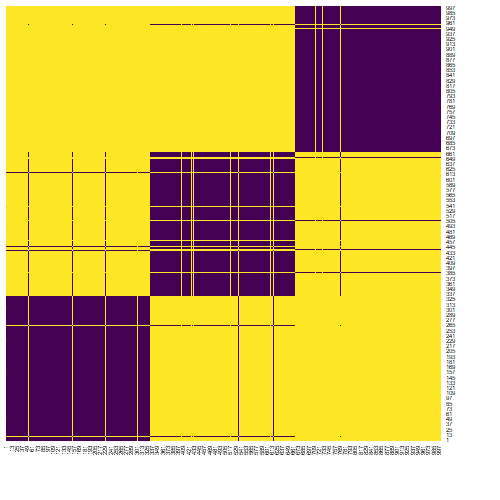}	
		\caption{Entropy regularized Binder loss clustering $\lambda = 10$.}
	\end{subfigure}	
 \hfill
\begin{subfigure}[t]{0.2\textwidth}
	\centering
 \includegraphics[width=\textwidth]{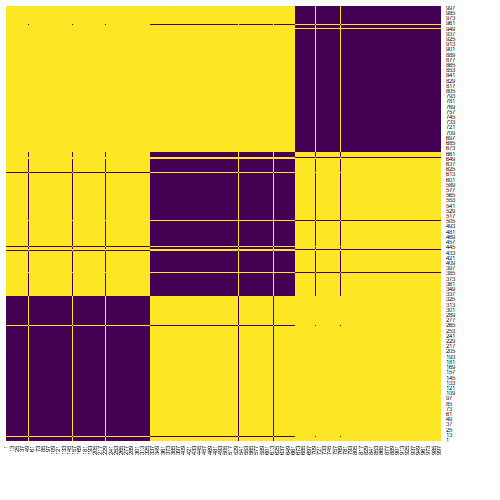}
	\caption{Entropy regularized Binder loss clustering $\lambda = 20$.}
\end{subfigure}	
	\caption{\label{fig:estsim_app} Gaussian simulation study ($\alpha$ random).
 Estimated clustering for the simulation study darker squares denote couples of observations clustered together. 
}
\end{figure}

\begin{figure}[H]
\vspace{0.2cm}
	\begin{subfigure}[b]{0.3\textwidth}
		\centering
  \includegraphics[width=\textwidth]{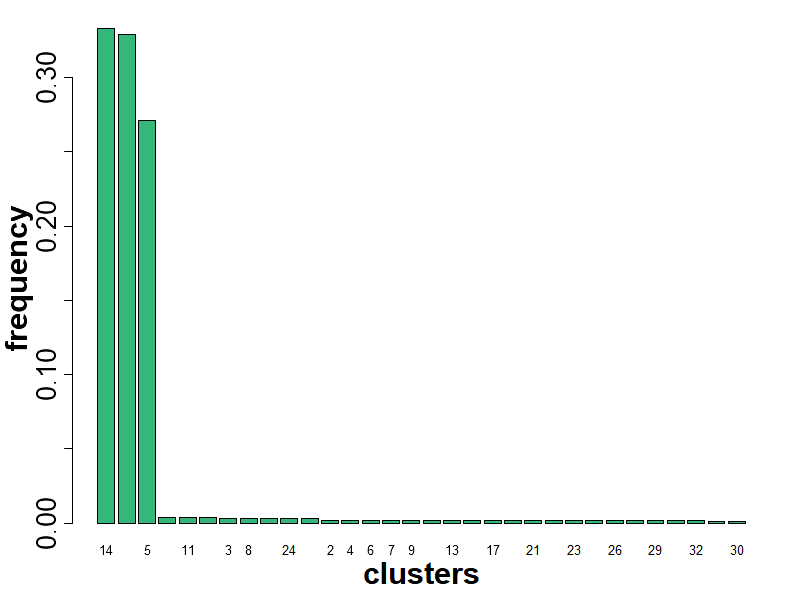}
		\caption{Without entropy regularization.}
	\end{subfigure}	
\hfill
	\begin{subfigure}[b]{0.3\textwidth}
		\centering
  \includegraphics[width=\textwidth]{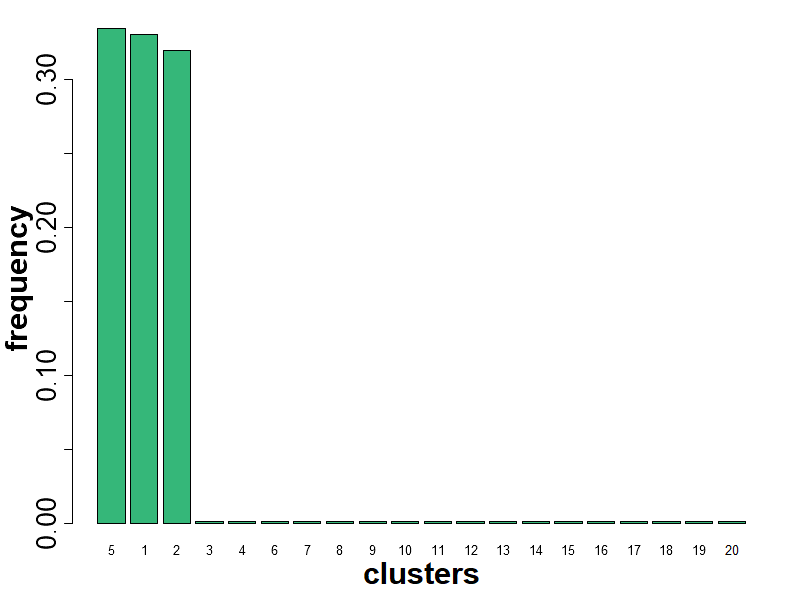}
		\caption{With entropy regularization for $\lambda = 10$.}
	\end{subfigure}	
\hfill
	\begin{subfigure}[b]{0.3\textwidth}
		\centering
  	\includegraphics[width=\textwidth]{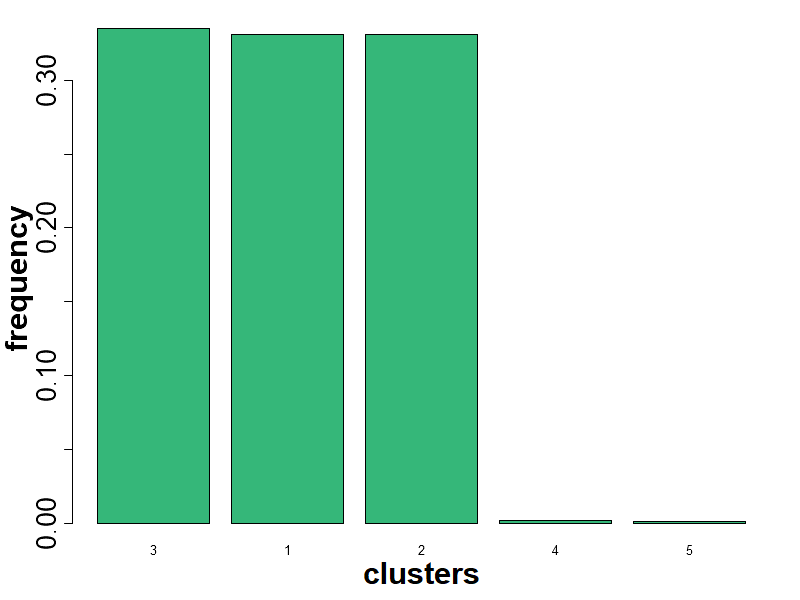}
		\caption{With entropy regularization for $\lambda = 20$.}
	\end{subfigure}	
	\caption{\label{fig:estfreq_app} Gaussian simulation study ($\alpha$ random).
 Estimated number of clusters and clusters' frequencies.}
\end{figure}

Recalling that we defined as sparsely populated clusters those clusters containing $10\%$ or less of observations, we found that $6 \, 040$ out $15\,000$) of the MCMC iterations, $10\%$ or more of the observations are allocated into sparsely populated clusters, while in $9 \, 908$ out of $15\,000$ of MCMC iterations, $5\%$ or more of the observations are allocated into sparsely populated clusters, see Figure~\ref{fig:hist_app}. 
The same counts after entropy-regularization of the posterior are, with $\lambda = 10$, 2772 and 5269 out 15 000, see Figure~\ref{fig:histreg10_app}, and, with $\lambda = 20$, 417 and 1302 out 15 000, see Figure~\ref{fig:histreg_app}. 
Figure~\ref{fig:estsim_app} shows the ground truth and the estimated clusters with and without entropy regularization. 
The number of observations misclassified (with respect to the simulation truth) with the Binder loss point estimate ($\lambda=0$) is $80$, with the regularization with $\lambda=10$ is $37$ and with $\lambda=20$ is $25$, showing an improvement also due to regularization also with this model.
Figure~\ref{fig:estfreq_app} shows the cluster frequencies for the three point-estimates.
Finally, we note that also with $\alpha$ random in this simple univariate Gaussian kernel simulation scenario we can obtain the correct number of occupied components (i.e., clusters) and thus avoid sparsely populated clusters in the point estimate using the more parsimonious VI loss (with the default parameter $a=1$ as implemented in \textsc{salso}) instead of considering the Binder loss (with the default parameter $a=1$).

\FloatBarrier
\bibliographystyle{natbib}
\bibliography{BNP_cit, Rebaudo_pub}

\end{document}